\documentclass[journal,10pt,letterpaper]{IEEEtran}
\usepackage{booktabs}
\usepackage{threeparttable}
\usepackage[justification=centering]{caption}
\IEEEoverridecommandlockouts    
\usepackage{subcaption}
\usepackage[utf8]{inputenc}
\usepackage[T1]{fontenc}
\usepackage{graphicx}
\usepackage{float}
\usepackage{threeparttable}
\usepackage{xcolor}
\usepackage[normalem]{ulem}
\graphicspath{{./graphic/}}
\usepackage{svg}
\usepackage{multirow}
\usepackage{algorithmicx}
\usepackage{algpseudocode}
\usepackage{algorithm}
\usepackage{amsmath}
\usepackage{amssymb}
\usepackage{cite}
\usepackage{hyperref}
\hypersetup{
    colorlinks=true,
    linkcolor=blue,
    citecolor=blue,
    filecolor=blue,
    urlcolor=blue
}
\newcommand{\custsection}[1]{
  \textsc{\large#1}
}
\makeatother
\title{\LARGE \bf Dual-Domain Deep Learning-Assisted NOMA-CSK Systems for Secure and Efficient Vehicular Communications}

\author{Tingting Huang, Jundong Chen, Huanqiang Zeng,~\IEEEmembership{Senior~Member,~IEEE}, Guofa Cai,~\IEEEmembership{Senior~Member,~IEEE} \\ and Georges Kaddoum,~\IEEEmembership{Senior~Member,~IEEE}
\thanks{ Corresponding author: Jundong Chen. Tingting Huang and Jundong Chen are with the School of Engineering, Huaqiao University, Quanzhou 362021, China (e-mail: JDchan@163.com; tthuang@hqu.edu.cn).}%
\thanks{Huanqiang Zeng is with the School of Information Science and Engineering, Huaqiao University, Xiamen 361021, China (e-mail: zeng0043@hqu.edu.cn).}
\thanks{Guofa Cai is with the School of Information Engineering, Guangdong University of Technology, Guangzhou 510006, China (e-mail: caiguofa2006@gdut.edu.cn).}
\thanks{Georges Kaddoum is with the Electrical Engineering Department, $\acute{\mathrm{E}}$cole de Technologie Sup$\acute{\mathrm{e}}$rieure, University of Quebec, Montreal, QC H2L 2C4, Canada, and also with the Artificial Intelligence \& Cyber Systems Research Center, Lebanese American University (e-mail: georges.kaddoum@etsmtl.ca).}
}

\begin{document}
\maketitle
\thispagestyle{empty}
\pagestyle{empty}
\begin{abstract}
Ensuring secure and efficient multi-user (MU) transmission is critical for vehicular communication systems. Chaos-based modulation schemes have garnered considerable interest due to their benefits in physical layer security. However, most existing MU chaotic communication systems, particularly those based on non-coherent detection, suffer from low spectral efficiency due to reference signal transmission, and limited user connectivity under orthogonal multiple access (OMA). While non-orthogonal schemes, such as sparse code multiple access (SCMA)-based DCSK, have been explored, they face high computational complexity and inflexible scalability due to their fixed codebook designs. This paper proposes a deep learning-assisted power domain non-orthogonal multiple access chaos shift keying (DL-NOMA-CSK) system for vehicular communications. A deep neural network (DNN)-based demodulator is designed to learn intrinsic chaotic signal characteristics during offline training, thereby eliminating the need for chaotic synchronization or reference signal transmission. The demodulator employs a dual-domain feature extraction architecture that jointly processes the time-domain and frequency-domain information of chaotic signals, enhancing feature learning under dynamic channels. The DNN is integrated into the successive interference cancellation (SIC) framework to mitigate error propagation issues. Theoretical analysis and extensive simulations demonstrate that the proposed system achieves superior performance in terms of spectral efficiency (SE), energy efficiency (EE), bit error rate (BER), security, and robustness, while maintaining lower computational complexity compared to traditional MU-DCSK and existing DL-aided schemes. These advantages validate its practical viability for secure vehicular communications.

\end{abstract}
\bigskip

\section{Introduction}

With the rapid advancement of autonomous driving technologies and the widespread deployment of intelligent transportation systems (ITS), vehicular communications have witnessed dramatic growth in demand \cite{1-1}. However, the frequent information exchange among vehicles, coupled with the inherent openness of wireless channels, exposes vehicular communication systems to potential security threats \cite{1-3}. 

Due to their noise-like properties and sensitivity to initial conditions, chaos-based secure communication schemes have attracted considerable attention in recent years \cite{1-4,1-5,1-6,1-7}. Such schemes can enhance physical layer security by exploiting the inherent randomness of chaotic signals \cite{1-7}. Furthermore, chaotic communication systems exhibit relatively low implementation complexity, making them particularly suitable for resource-constrained vehicular environments \cite{1-8}.

In chaotic communication  systems, detection schemes are commonly categorized into coherent and non-coherent approaches \cite{1-9}. Coherent schemes, such as chaos shift keying (CSK) systems \cite{CSK.1,CSK.2}, information is modulated onto chaotic signals, and the receiver must synchronize and reconstruct identical copies of these chaotic signals for demodulation. However, achieving reliable chaotic synchronization is challenging in practical environments due to the sensitivity to initial conditions and channel impairments \cite{sync-challenge}.

Most of the existing chaos-based communication systems are based on non-coherent schemes, such as differential chaos shift keying (DCSK) \cite{DCSK}. In DCSK, a copy of the chaotic signal is transmitted as a reference signal along with the information-bearing signal over the channel, thereby eliminating the need for complex chaotic synchronization at the receiver. However, in multi-user (MU) scenarios, conventional DCSK schemes suffer from severe multi-user interference (MUI), and require the transmission of substantial reference signals, thereby significantly reducing spectral efficiency and system performance \cite{1-10}.


To address these limitations, many variants of the MU-DCSK system have been proposed \cite{MU-DCSK.1,MU-DCSK.2,MU-DCSK.3,MU-DCSK.4,MU-DCSK.5}. Specifically, the authors in \cite{MU-DCSK.1} proposed a MU orthogonal frequency division multiplexing (OFDM) DCSK system, where each user utilizes dedicated subcarriers for reference signal transmission and shares the remaining subcarriers with other users to transmit information symbols, enabling the system to transmit $M$ bits using only $N_P$ chaotic reference signals $(N_P \ll M)$, thereby increasing spectral efficiency. Furthermore, the authors in \cite{MU-DCSK.2} proposed a multicarrier $M$-ary orthogonal chaotic vector shift keying with index modulation (MC-MOCVSK-IM) system, where information bits are conveyed not only by multiple groups of $M$-ary information-bearing signals, but also by the specific indices of the selected reference signals, thereby achieving improved energy and spectral efficiency.

To enhance security and robustness, a frequency-time diversity-aided OFDM-DCSK system was proposed in \cite{MU-DCSK.3}, which employs non-repetitive frequency hopping operations on both reference chips and chaotic modulated symbols across different subcarriers. Additionally, to address the MUI problem in MU-DCSK systems, a novel receiver design was proposed in \cite{MU-DCSK.4} that exploits the low-rank structure of the received signal matrix through a least-squares-based optimization framework. Furthermore, a joint differential pulse position modulation (DPPM) and DCSK with Walsh codes (DPPM-DCSK-WC) system was proposed in \cite{MU-DCSK.5}, which employs modified Walsh codes to mitigate energy and rate waste.

However, most of these improved MU-DCSK schemes still rely on orthogonal multiple access (OMA), which limits their ability to support massive connectivity in future vehicular networks. Non-orthogonal multiple access (NOMA), as a key technology for next-generation communication systems, has emerged as a promising solution for achieving massive connectivity \cite{1-11,1-12}. Representative NOMA techniques include power domain NOMA (PD-NOMA) \cite{PD-NOMA}, sparse code multiple access (SCMA) \cite{SCMA}, and pattern division multiple access (PDMA) \cite{PDMA}. By allowing multiple users to share the same time-frequency resources, NOMA can achieve superior spectral efficiency, massive connectivity, and lower transmission latency compared to OMA schemes \cite{1-11}.

To leverage the advantages of NOMA, existing studies have integrated SCMA with DCSK systems \cite{SCMA-DCSK.1,SCMA-DCSK.2}. However, SCMA-DCSK systems face significant challenges, including high computational complexity due to message passing algorithms (MPA) at the receiver and limited user scalability due to fixed codebook designs. Although recent studies \cite{DL-SCMA-DCSK} have employed deep learning (DL) techniques to reduce receiver complexity in SCMA-DCSK systems, these schemes still require fixed codebook designs and reference signal transmission, thereby limiting their flexibility and spectral efficiency.

In contrast, PD-NOMA employs power domain superposition coding, which can flexibly support different numbers of users without requiring fixed codebook designs. Moreover, PD-NOMA utilizes successive interference cancellation (SIC) \cite{b5} for signal detection at the receiver, which significantly reduces computational complexity compared to the MPA-based detection in SCMA systems \cite{SCMA_and_PD-NOMA}.

Motivated by these observations and recent advances in DL that eliminate the need for chaotic synchronization in chaotic communication systems \cite{DLCSK,hyper-parameter-2,r3-2}, this paper proposes a DL-assisted power domain NOMA CSK (DL-NOMA-CSK) system. The proposed system aims to achieve spectral-efficient and low-complexity MU transmission while preserving the security advantages of chaotic communications in vehicular networks. Specifically, a deep neural network (DNN)-based demodulator is employed to optimize the SIC detection process, effectively mitigating error propagation and error floor issues in conventional SIC receivers \cite{DL-NOMA}. 


In contrast to traditional MU-DCSK schemes \cite{MU-DCSK.1,MU-DCSK.2,MU-DCSK.3,MU-DCSK.4,MU-DCSK.5} that require reference signal transmission, the proposed DNN demodulator is trained to learn the intrinsic characteristics of chaotic signals during the offline training phase. This allows the system to directly demodulate chaotic signals in the online deployment without requiring chaotic synchronization or reference signal transmission, thereby significantly improving spectral efficiency and reducing system complexity.

Moreover, compared to existing DL-aided chaotic communication schemes \cite{DL-SCMA-DCSK,DL-NOMA-DCSK.1,DL-NOMA-DCSK.2,DL-NOMA-DCSK.3}, the proposed DNN architecture employs a dual-domain feature extraction mechanism that jointly processes temporal and spectral characteristics of chaotic signals. This complementary representation enables the demodulator to extract richer discriminative features, thereby improving the bit error rate (BER) performance under dynamic vehicular channel conditions.

The main contributions of this paper are as follows:

\begin{enumerate}
\item We propose a DL-assisted NOMA-CSK transceiver system where a DNN-based demodulator is trained to learn the intrinsic characteristics of chaotic signals, eliminating the need for chaotic synchronization or reference signal transmission and significantly improving spectral efficiency while reducing transmitter complexity.

\item We design a dual-domain DNN-based demodulator architecture that exploits the intrinsic characteristics of chaotic signals to jointly process their time-domain and frequency-domain information, enabling effective feature learning and enhanced BER performance under dynamic vehicular channel conditions.

\item We develop a DNN-enhanced SIC framework that effectively addresses error propagation issues in NOMA detection. Performance analysis and simulations demonstrate superior performance in BER, spectral efficiency, computational complexity, security, and robustness compared to traditional MU-DCSK and other DL-aided schemes.

\end{enumerate}

The remainder of this paper is organized as follows. Section \ref{2} presents the transceiver structure of the proposed DL-NOMA-CSK system. Section \ref{DNN} describes the DNN-based demodulator architecture, including the operating principles of each layer, the offline training process, and hyperparameter selection criteria. 
Section \ref{4} provides a performance analysis of complexity, energy efficiency (EE), spectral efficiency (SE), and security. Section \ref{Simulation Results and Discussions} discusses dataset generation and presents comprehensive simulation results, including BER performance, security, and robustness.
Finally, Section \ref{6} concludes this paper.

\section{System Model}
\label{2}

\begin{figure}[t] 
		\centering
        \captionsetup{justification=raggedright,labelsep=period,singlelinecheck=off} 
		\includegraphics[width=\columnwidth]{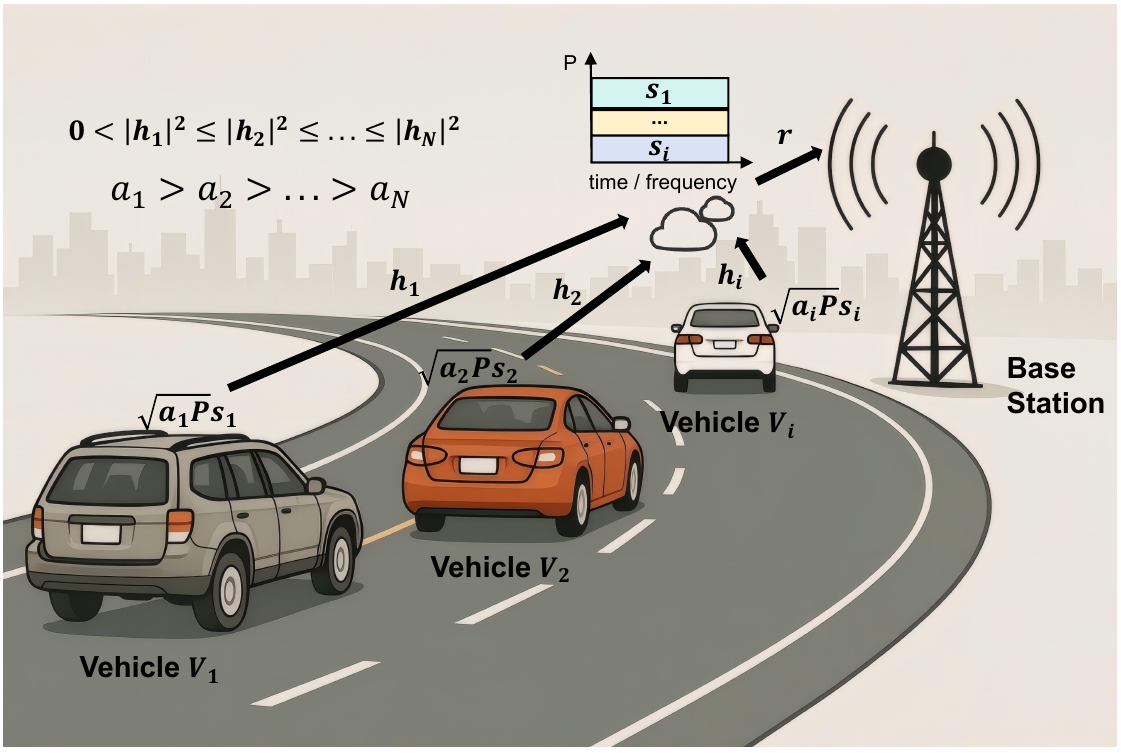} 
		\caption{Uplink V2I communication scenario with the proposed DL-NOMA-CSK scheme.}
		\label{Uplink V2I NOMA-DCSK communication scenario}
	\end{figure}

Consider an uplink vehicular cellular communication scenario comprising one base station (BS) and $N$ vehicles $V_i$, $i \in \{1, 2, \ldots, N\}$, as illustrated in Fig.~\ref{Uplink V2I NOMA-DCSK communication scenario}. These vehicles simultaneously transmit information to the BS using identical time-frequency resources with different transmit powers. Each node is equipped with a single antenna. Note that a similar downlink scenario can also be considered, where the proposed DL-NOMA-CSK scheme would be equally applicable.

\subsection{The Transmitter Structure}    

\begin{figure*}[t] 
		\centering
        \captionsetup{justification=raggedright,labelsep=period,singlelinecheck=off} 
		\includegraphics[width=\textwidth]{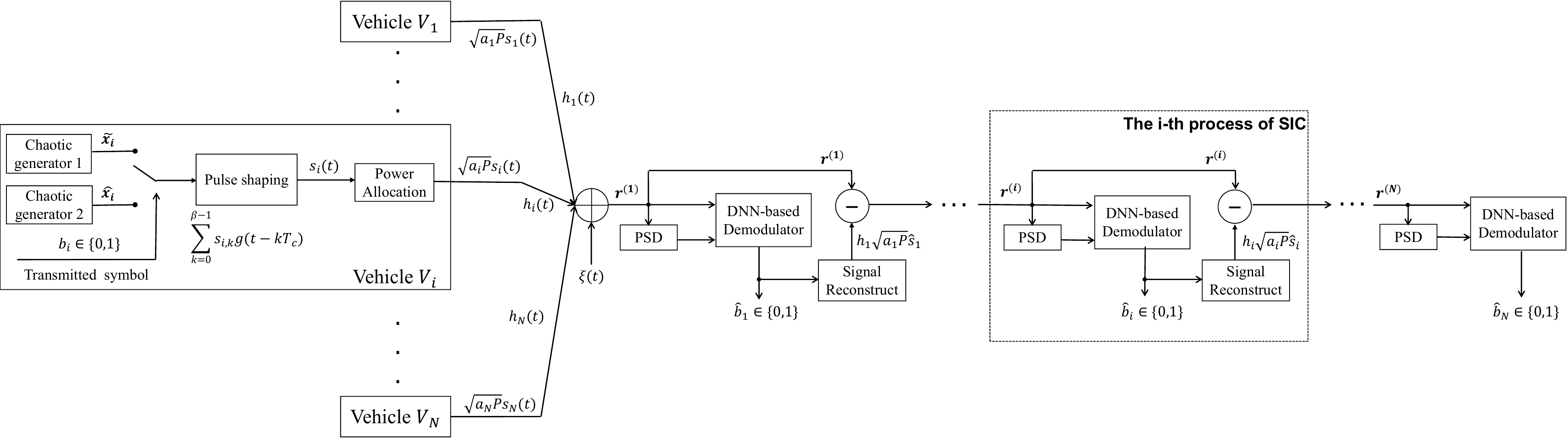} 
		\caption{The transceiver structure of the DL-NOMA-CSK system.}
		\label{Transceiver structure of the DL-NOMA-DCSK system.}
	\end{figure*}

Fig.~\ref{Transceiver structure of the DL-NOMA-DCSK system.} shows the proposed DL-NOMA-CSK transceiver structure. For vehicle $V_i$, the chaos generators employ Logistic map $\boldsymbol{\tilde{x}}_i$ and Cubic map $\boldsymbol{\widehat{x}}_i$, selected for their complementary statistical properties \cite{r3-2}, expressed as:
\begin{subequations}
\begin{align}
\tilde{x}_{i,k} = 3.7\tilde{x}_{i,k-1}(1-\tilde{x}_{i,k-1}),  \quad 0 \leq k \leq \beta-1,\\
\widehat{x}_{i,k} = 4\widehat{x}_{i,k-1}^3 - 3\widehat{x}_{i,k-1}, \quad 0 \leq k \leq \beta-1,
\end{align}
\end{subequations}
where $\tilde{x}_{i,k}$ and $\widehat{x}_{i,k}$ denotes the $k$-th chip of chaotic sequence $\boldsymbol{\tilde{x}}_i$ and $\boldsymbol{\widehat{x}}_i$, respectively, and the spreading factor $\beta$ is defined as the number of chaotic samples. Each transmitted binary symbol has a duration of $T_b$, while each chaotic chip has a duration of $T_c$. Consequently, each bit duration contains $\beta$ chaotic chips, i.e., $T_b = \beta T_c$.

Let $b_i \in \{0, 1\}$ represent the transmitted binary symbol for the $i$-th vehicle. For each bit transmission, the initial conditions $\tilde{x}_{i,0}, \widehat{x}_{i,0} \in (0, 1)$ of the chaotic sequences are randomly generated. During the symbol transmission duration, the spreading signal $s_i(t)$ for vehicle $V_i$ is expressed as:
\begin{equation}
s_i(t) = \sum_{k=0}^{\beta-1} s_{i,k} g(t - kT_c), \quad 0 \leq t \leq T_b,
\label{map}
\end{equation}
where $s_{i,k}$ denotes the $k$-th chaotic chip for vehicle $V_i$, defined as:
\begin{equation}
s_{i,k} = \begin{cases}
\tilde{x}_{i,k} & \text{if } b_i = 0 \\
\widehat{x}_{i,k} & \text{if } b_i = 1,
\end{cases}
\end{equation}
and $g(t)$ represents the chip pulse shaping. Unlike conventional chaotic  communications systems, the proposed scheme eliminates the need for reference signal transmission, thereby removing the requirement for delay lines in the transmitter structure.

We assume that wireless links between all vehicles and the BS experience independent and identically distributed (i.i.d.) block fading, where $h_i$ represents the channel coefficient between $V_i$ and the BS, which remains constant within each transmission block \cite{fixed}. Without loss of generality, the vehicles are ordered according to their channel gains as $0 < |h_1|^2 \leq |h_2|^2 \leq \cdots \leq |h_N|^2$, such that vehicle $V_1$ experiences the weakest channel condition.

Next, each vehicle is allocated transmit power according to the uplink PD-NOMA principle \cite{PD-NOMA}, where vehicles with weaker channel conditions are assigned higher power coefficients, such that $\alpha_1 > \alpha_2 > \cdots > \alpha_N$. Specifically, the power allocation coefficient for the $i$-th vehicle is given by:
\begin{equation}
\alpha_i = \frac{2^{N-i}}{\sum_{j=1}^{N} 2^{N-j}},
\label{eq:power_allocation}
\end{equation}
where the coefficients satisfy the constraint $\sum_{i=1}^{N} \alpha_i = 1$ with $0 < \alpha_i < 1$ for all $i \in \{1, 2, \ldots, N\}$. In practice, the BS periodically estimates the channel gains through pilot signals and broadcasts the corresponding power allocation coefficients to all vehicles \cite{1-11}.

With the power allocation coefficients determined, the transmitted signal for vehicle $V_i$ is expressed as:
\begin{equation}
\tilde{s}_i(t) = \sqrt{\alpha_i P}s_i(t), \quad 0 \leq t \leq T_b,
\label{eq:transmitted_signal}
\end{equation}
where $P$ represents the reference transmit power for each vehicle.
    
\subsection{The Receiver Structure}
    
All vehicles transmit information simultaneously over the same time-frequency resources, which are superimposed at the receiver. Under multipath Rayleigh fading \cite{r3-2}, the channel model of vehicle $V_i$ is given by:
\begin{equation}
\label{rayleigh}
h_i(t) = \sum_{l=1}^{L_i}\lambda_{i,l}\delta(t-\tau_{i,l}), \quad 0 \leq t \leq T_b,
\end{equation}
where $L_{i}$ represents the number of multipath components, $\lambda_{i,l}$ and $\tau_{i,l}$ are complex path gain and time delay of the $l$-th path, respectively. For each path $l$ of vehicle $V_i$, $\lambda_{i,l} = \psi_{i,l,1} + j\psi_{i,l,2}$, where $\psi_{i,l,1}$ and $\psi_{i,l,2}$ are independent Gaussian random variables with zero mean and variance $1/(2L_i)$.

Consequently, the received signal is expressed as:
\begin{equation}
r(t) = \sum_{i=1}^{N}h_i(t) \otimes \tilde{s}_i(t) + \xi(t), \quad 0 \leq t \leq T_b,
\end{equation}
where $\otimes$ denotes the convolution operator, and $\xi(t)$ is the additive white Gaussian noise (AWGN) with zero mean and power spectral density $N_0/2$.

Subsequently, the received signal $r(t)$ is processed by matched filtering and sampled every $T_c$ to obtain the discrete sequence $\boldsymbol{r} = [r_0, r_1, \ldots, r_{\beta-1}]$.

To decode the superimposed signals, DNN-based demodulation is employed to optimize the SIC detection process. The SIC process exploits the power disparity among user signals to sequentially decode and eliminate interference. The decoding procedure as follows:

1) First, for vehicle $V_1$ with the highest power allocation, the DNN-based demodulator $\mathcal{D}(\cdot)$: $\mathbb{R}^{2 \times \beta} \rightarrow \{0, 1\}$ processes the received signal, i.e., $\boldsymbol{r}^{(1)} = \boldsymbol{r}$, to decode the transmitted symbol $\hat{b}_1$, while treating all other user signals as interference.

2) After successfully decoding $\hat{b}_1$, the corresponding signal $\boldsymbol{\hat{s}}_1$ is reconstructed by using the same chaotic map as determined by $\hat{b}_1$, i.e., Logistic map if $\hat{b}_1=0$ and Cubic map if $\hat{b}_1=1$, with a randomly generated initial condition. The reconstructed signal is then subtracted from $\boldsymbol{r}^{(1)}$ to obtain the residual signal $\boldsymbol{r}^{(2)} = \boldsymbol{r}^{(1)} - \boldsymbol{h}_1\sqrt{\alpha_1 P} \boldsymbol{\hat{s}}_1$. This residual signal is then processed by $\mathcal{D}(\cdot)$ to decode $\hat{b}_2$.

3) The process continues iteratively for the remaining vehicles $V_i$ ($i = 3, 4, \ldots, N$). For each iteration, the DNN demodulator $\mathcal{D}(\cdot)$ processes the current residual signal $\boldsymbol{r}^{(i)}$ to decode $\hat{b}_i$. After decoding, $\boldsymbol{\hat{s}}_i$ is reconstructed following the same procedure as in step 2), and the residual signal is updated as $\boldsymbol{r}^{(i+1)} = \boldsymbol{r}^{(i)} - \boldsymbol{h}_i\sqrt{\alpha_i P} \boldsymbol{\hat{s}}_i$. This process continues until all $N$ vehicles are decoded.

Note that in practice, the system would include a channel estimation unit, typically employing pilot-based methods \cite{CSI.1,CSI.2}. However, for model simplicity and to focus on the core contribution of the proposed DL-NOMA-CSK scheme, we assume perfect channel state information (CSI) in the theoretical analysis \cite{b6}. The impact of imperfect CSI on system performance will be evaluated through simulations in Section \ref{Robustness Analysis}.

\begin{algorithm}
\hrule
\caption{SIC Process with DNN-based Demodulator}
\label{algorithm1}
\hrule
\vspace{0.2em}
\textbf{Input:} Composite received signal ${r}(t)$, power allocation coefficients $\{\alpha_i\}_{i=1}^N$, channel coefficients $\{\boldsymbol{h}_i\}_{i=1}^N$ 
\vspace{0.2em}

\textbf{Output:} Estimated transmitted bits $\hat{\boldsymbol{b}} = \{\hat{b}_i\}_{i=1}^N$
\begin{algorithmic}[1]
\State \textbf{Initialization:} $\boldsymbol{r}^{(1)} \leftarrow \boldsymbol{r}$
\For{$i = 1$ to $N$}
    \State Construct feature tensor: $\boldsymbol{F}^{(i)} \leftarrow \begin{bmatrix} \boldsymbol{r}^{(i)} \\ \boldsymbol{S}_{\boldsymbol{r}^{(i)}} \end{bmatrix}$
    \State $\hat{b}_i \leftarrow \mathcal{D}(\boldsymbol{F}^{(i)})$ 
    \If{$i < N$}
        \State Generate random initial condition $x_{i,0} \sim \mathcal{U}(0,1)$
        \State $\boldsymbol{\hat{s}}_i \leftarrow \begin{cases} \text{Logistic}(x_{i,0}) & \hat{b}_i = 0 \\ \text{Cubic}(x_{i,0}) & \hat{b}_i = 1 \end{cases}$
        \State $\boldsymbol{r}^{(i+1)} \leftarrow \boldsymbol{r}^{(i)} - \boldsymbol{h}_i\sqrt{\alpha_i P}\boldsymbol{\hat{s}}_i$
    \EndIf
\EndFor
\State \textbf{Return} $\hat{\boldsymbol{b}} = [\hat{b}_1, \hat{b}_2, \ldots, \hat{b}_N]$
\end{algorithmic}
\hrule
\end{algorithm}

At each stage of the SIC process, the signal processing involves dual-domain feature extraction to capture both temporal and spectral characteristics. Specifically, the time-domain features consist of the raw residual signal samples $\boldsymbol{r}^{(i)}$. For spectral characteristics, we compute the power spectral density (PSD) estimate as follows:
\begin{equation}
\boldsymbol{S}_{\boldsymbol{r}^{(i)}}[k] = \left|\mathcal{F}\{\boldsymbol{r}^{(i)}\}[k]\right|^2, \quad k = 0, 1, \ldots, \beta-1,
\end{equation}
where $\mathcal{F}\{\cdot\}$ denotes the discrete Fourier transform (DFT). 

The resulting dual-domain feature tensor can be formulated as $\boldsymbol{F}^{(i)} \in \mathbb{R}^{2 \times \beta}$, where
\begin{equation}
\boldsymbol{F}^{(i)} = \begin{bmatrix} \boldsymbol{r}^{(i)} \\ \boldsymbol{S}_{\boldsymbol{r}^{(i)}} \end{bmatrix}
\end{equation}

This 2D tensor feeds the DNN-based demodulator $\mathcal{D}(\cdot)$: $\mathbb{R}^{2 \times \beta} \rightarrow \{0, 1\}$, which maps features to binary decisions for $\hat{b}_i$ decoding. Algorithm~\ref{algorithm1} summarizes the SIC process with DNN-based demodulation, and the architecture and training methodology of the DNN-based demodulator detailed in Section~\ref{DNN}.

Notably, the proposed DL-NOMA-CSK system effectively mitigates the error propagation and imperfect interference cancellation inherent in conventional SIC-based receivers \cite{b7}. By integrating time-domain signal samples and frequency-domain PSD estimates into a unified feature tensor, the dual-domain approach provides complementary signal representations that enhance detection robustness.

\section{DNN-based Demodulator}
\label{DNN}
\begin{figure*}[!t] 
		\centering
        \captionsetup{justification=raggedright,labelsep=period,singlelinecheck=off} 
		\includegraphics[width=0.9\textwidth]{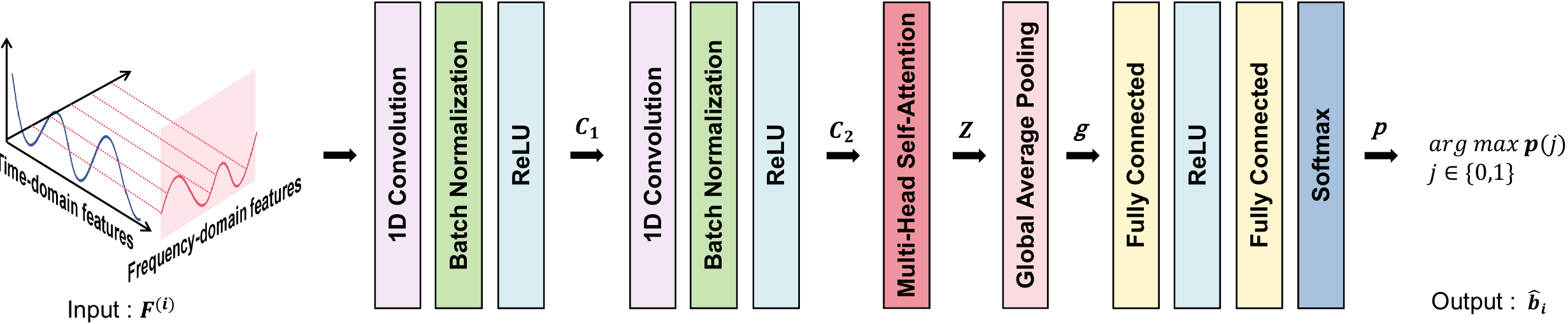} 
		\caption{The architecture of DNN-based demodulator.}
		\label{Architecture of DNN-based demodulator}
\end{figure*}

Fig.~\ref{Architecture of DNN-based demodulator} illustrates the architecture of the proposed DNN-based demodulator, which consists of two one-dimensional convolutional (1D Conv) layers, two batch normalization (BN) layers, a multi-head self-attention (MH-SA) layer, a global average pooling (GAP) layer, two fully-connected layers (FCLs), and regularization components. This architecture is designed to leverage hierarchical feature extraction for effectively capturing the complex temporal dependencies and spectral characteristics inherent in chaotic signals. 
The dual-domain input provides complementary information that improving the demodulator's ability to decode chaotic modulated symbols under interference.


\subsection{Operation Principles of DNN Layers}

\subsubsection{Hierarchical Convolutional Processing}

In this paper, we employ two convolutional modules, each comprising a 1D Conv layer and a BN layer, to extract discriminative features from the dual-domain input tensor $\boldsymbol{F}^{(i)}$. 

Specifically, the first convolutional module applies to the input tensor $\boldsymbol{F}^{(i)} \in \mathbb{R}^{2 \times \beta}$, where each convolutional kernel processes adjacent elements along the feature dimension to extract discriminative signal representations from both time and frequency domains. The operation can be expressed as:
\begin{equation}
\boldsymbol{z}_1(m,j) = \sum_{d=0}^{1}\sum_{l=0}^{k_{size}-1} \boldsymbol{W}_1(m,d,l) \cdot \boldsymbol{F}^{(i)}(d,j+l) + \boldsymbol{b}_1(m),
\end{equation}

where $\boldsymbol{W}_1 \in \mathbb{R}^{n \times 2 \times k_{size}}$ represents the learnable weight tensor, $\boldsymbol{b}_1 \in \mathbb{R}^{n}$ denotes the corresponding bias vector, $d\in \{0,1\}$ indexes the dual-domain input channels, $m \in \{0,1,\ldots,n-1\}$ indexes the filters, and $j \in \{0,1,\ldots,\beta - k_{size} + 1\}$ indexes the output feature positions. The convolution operation slides the kernel across the feature dimension of the input tensor with unit stride, producing an output tensor $\boldsymbol{z}_1 \in \mathbb{R}^{n \times (\beta-k_{size}+1)}$. 
The parameter $n$ denotes the number of convolutional filters, determining the dimensionality of the feature maps produced, while $k_{size}$ represents the convolutional kernel size that defines the receptive field of each filter. 
The convolution output $\boldsymbol{z}_1$ is subsequently processed through BN to mitigate internal covariate shift, followed by ReLU activation:
\begin{equation}
\boldsymbol{C}_1 = \text{ReLU}(\text{BN}(\boldsymbol{z}_1)),
\end{equation}
where $\text{BN}(\cdot)$ represents the BN operation that standardizes the intermediate features, and ReLU introduces non-linearity defined as $\text{ReLU}(x) = \max(0,x)$. 

Subsequently, the second convolutional module processes the feature maps $\boldsymbol{C}_1 \in \mathbb{R}^{n \times (\beta-k_{size}+1)}$ from the first module for enhanced feature representation. Specifically, this module employs an expanded convolution kernel size of $2k_{size}$ to extend the receptive field, thereby capturing longer-range dependencies inherent in chaotic sequences. The convolution operation can be formulated as:
\begin{equation}
\boldsymbol{z}_2(m,j) = \sum_{d=0}^{n-1}\sum_{l=0}^{2k_{size}-1} \boldsymbol{W}_2(m,d,l) \cdot \boldsymbol{C}_1(d,j+l) + \boldsymbol{b}_2(m),
\end{equation}
where $\boldsymbol{W}_2 \in \mathbb{R}^{n \times n \times 2k_{size}}$ represents the learnable weight tensor of the second module, $\boldsymbol{b}_2 \in \mathbb{R}^{n}$ denotes the corresponding bias vector, and $j \in \{0,1,\ldots,(\beta-k_{size}+1) - 2k_{size} + 1\}$ indexes the output positions. Similar to the first module, the output is processed through BN and ReLU activation:
\begin{equation}
\boldsymbol{C}_2 = \text{ReLU}(\text{BN}(\boldsymbol{z}_2)),
\end{equation}
yielding the output feature maps $\boldsymbol{C}_2 \in \mathbb{R}^{n \times L}$, where $L = (\beta-k_{size}+1) - (2k_{size}-1) = \beta - 3k_{size} + 2$ represents the temporal dimension of the features after two convolutional stages.

This hierarchical configuration enables the progressive abstraction of signal characteristics, with the first module extracting fundamental signal features from the dual-domain input and the second module identifying more complex feature structures through its expanded receptive field. The resulting feature maps $\boldsymbol{C}_2$ provide rich feature representations for subsequent network stages, enabling robust demodulation across varying channel conditions.

\subsubsection{Multi-Head Self-Attention Mechanism}

Following the hierarchical convolutional processing, an MH-SA layer is incorporated to dynamically model long-range dependencies and capture global contextual information from the extracted feature maps $\boldsymbol{C}_2$. 
This capability is particularly crucial for chaotic signals that exhibit complex nonlinear dynamics and long-range temporal dependencies.

\begin{figure}[!t] 
		\centering
        \captionsetup{justification=raggedright,labelsep=period,singlelinecheck=off} 
		\includegraphics[width=0.8\columnwidth]{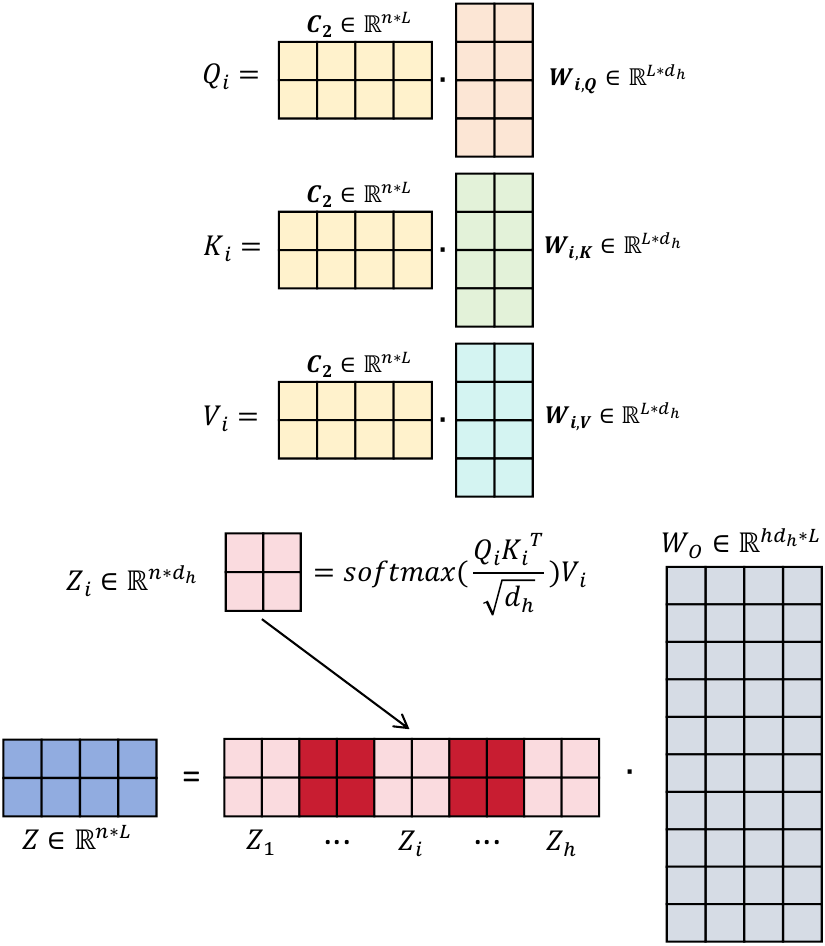} 
		\caption{The architecture of the multi-head self-attention mechanism.}
		\label{computational process of attention}
	\end{figure}

As shown in Fig.~\ref{computational process of attention}, the implementation of MH-SA follows the scaled dot-product attention framework. Firstly, for each attention head $i \in \{1,2,\ldots,h\}$, we project the input features into three distinct linear projections: Query ($\mathbf{Q}$), Key ($\mathbf{K}$), and Value ($\mathbf{V}$) through learnable linear transformations, given by:
\begin{subequations}
\begin{align}
\mathbf{Q}_i &= \boldsymbol{C}_2 \cdot \mathbf{W}_{i,Q},\\
\mathbf{K}_i &= \boldsymbol{C}_2 \cdot \mathbf{W}_{i,K},\\
\mathbf{V}_i &= \boldsymbol{C}_2 \cdot \mathbf{W}_{i,V}.
\end{align}
\end{subequations}
where $\mathbf{W}_{i,Q}$, $\mathbf{W}_{i,K}$ and $\mathbf{W}_{i,V}\in \mathbb{R}^{L \times d_h}$ are learnable weight matrices for the $i$-th attention head, yielding projections $\mathbf{Q}_i, \mathbf{K}_i, \mathbf{V}_i \in \mathbb{R}^{n \times d_h}$, with $d_h$ representing the dimension of each attention head. 

Subsequently, for each attention head $i$, we compute the scaled dot-product attention to generate the feature representation $\mathbf{Z}_i$ as follows:
\begin{equation}
\mathbf{Z}_i = \text{softmax}\left(\frac{\mathbf{Q}_i \mathbf{K}_i^T}{\sqrt{d_h}}\right) \cdot \mathbf{V}_i,
\end{equation}

The softmax operation normalizes the attention scores into a probability distribution, ensuring that attention weights sum to unity for each feature position. The scaling factor $\sqrt{d_h}$ is employed to stabilize gradients during backpropagation, particularly when $d_h$ is large.

Finally, the outputs from all attention heads $\mathbf{Z}_i \in \mathbb{R}^{n \times d_h}$ are concatenated along the feature dimension and linearly transformed to produce the final attention output:
\begin{equation}
\mathbf{Z} = [\mathbf{Z}_1, \mathbf{Z}_2, \ldots, \mathbf{Z}_h] \cdot \mathbf{W}_O,
\end{equation}
where $\mathbf{W}_O \in \mathbb{R}^{(h \cdot d_h) \times L}$ is a learnable output projection matrix that maps the concatenated features back to the original dimension, yielding $\mathbf{Z} \in \mathbb{R}^{n \times L}$. 

This multi-head design enables the model to jointly capture signal features from different representation subspaces, enhancing its modeling capability for complex nonlinear chaotic signals and thereby improving the robustness of the demodulation process.

\subsubsection{Global Feature Aggregation}

Following the MH-SA mechanism, a GAP layer is employed to aggregate the features $\mathbf{Z} \in \mathbb{R}^{n \times L}$ into a compact representation, producing a feature vector $\boldsymbol{G}$. Mathematically, this operation can be expressed as:
\begin{equation}
\boldsymbol{G}(m) = \frac{1}{L}\sum_{j=0}^{L-1} \mathbf{Z}(m,j)
\end{equation}
where $m \in \{0,1,\ldots,n-1\}$ indexes the feature channels. The resulting feature vector $\boldsymbol{G} \in \mathbb{R}^{n}$ encapsulates the essential characteristics of the chaotic signal while abstracting away temporal details.

The GAP layer significantly reduces model complexity by eliminating flattening operations. Furthermore, it introduces temporal shift invariance, enabling the network to generalize effectively across varying temporal shifts in the received signal, which is particularly beneficial for chaotic communication systems operating in dynamic channel environments.

\subsubsection{Classification Decision}

The final stage of the DNN-based demodulator consists of a classification module, which transforms the global feature vector $\boldsymbol{G}$ obtained from the GAP layer into a binary decision corresponding to the transmitted bit. 

Specifically, this module employs two FCLs with nonlinear activation functions to perform the classification task. The first FCL projects the global features into a higher-dimensional representation space and applies the ReLU activation function to introduce nonlinearity, thereby enhancing the model's capability to establish complex decision boundaries.

Subsequently, the second FCL maps this intermediate representation to a two-dimensional output space $\boldsymbol{f} \in \mathbb{R}^{2}$, which corresponds to the unnormalized scores for the two possible transmitted bits. This output is then transformed into a probability distribution through the softmax function, expressed as:
\begin{equation}
\boldsymbol{p}(j) = \frac{e^{\boldsymbol{f}(j)}}{\sum_{k=0}^{1} e^{\boldsymbol{f}(k)}}, \quad j \in \{0,1\},
\end{equation}
where $\boldsymbol{p}(j)$ represents the probability that the transmitted bit equals $j$. 

Finally, the estimated transmitted bit $\hat{b}_i$ for the $i$-th vehicle at the current SIC stage is determined through a maximum a posteriori (MAP) decision rule:
\begin{equation}
\hat{b}_i = \underset{j \in \{0,1\}}{\arg\max} \ \boldsymbol{p}(j).
\end{equation}

This differentiable decision pipeline enables end-to-end optimization of the entire DNN-based demodulator, allowing the convolutional and attention layers to adapt their feature extraction capabilities to enhance classification performance. The proposed architecture demonstrates superior robustness against channel impairments and interference, particularly when integrated within the SIC framework of the DL-NOMA-CSK system.

\subsection{Offline Training of DNN}

During the offline training stage, the network's learnable parameters are updated through iterative training on the training dataset. Specifically, let $\mathcal{T} = \{(\boldsymbol{F}^{(i)}_j, b_j)\}$ represent the training dataset, where $\boldsymbol{F}^{(i)}_j \in \mathbb{R}^{2 \times \beta}$ denotes the dual-domain feature tensor at the $i$-th SIC stage for the $j$-th training sample, with $i \in \{1,2,\ldots,N\}$ representing different vehicles $V_i$, and $b_j \in \{0, 1\}$ corresponding to the transmitted symbol. During training, the optimization algorithm continuously adjusts the learnable parameters by minimizing the cross-entropy loss between the predicted labels $\hat{b}_j$ and ground-truth labels $b_j$ to achieve optimal convergence. Once trained, the optimized network parameters are fixed and deployed for online operation within the DL-NOMA-CSK system.

The training samples are constructed through Monte Carlo simulations incorporating various operational conditions typical of wireless communication scenarios. This approach ensures that the resulting demodulator exhibits robust generalization capabilities in practical deployments. More details about dataset generation are discussed in Section \ref{Dataset Discussion}.

\subsection{Hyperparameter Selection}
\label{3}

The efficacy of the proposed DNN-based demodulator is significantly influenced by judicious selection of various hyperparameters. This subsection presents several critical hyperparameters and their impact on system performance.

\subsubsection{Output Dimensions and Learnable Parameters}

Table~\ref{tab:dnn_parameters} presents the output dimensions and the number of learnable parameters for each layer in the proposed DNN architecture. The architectural complexity is primarily characterized by two fundamental parameters: the number of convolutional filters and the MH-SA mechanism parameters. In this study, we implement $n=32$ filters with kernel size $k_{size}=3$ and $h=8$ attention heads with per-head dimension $d_h=64$. This configuration was determined through ablation studies evaluating the trade-off between computational complexity and demodulation performance \cite{r3-2,hyper-parameter-1}. 

\begin{table}[t]
\centering
\caption{\custsection{Output Dimensions and Learnable Parameters in the Proposed DNN}}
\label{tab:dnn_parameters}
\renewcommand{\arraystretch}{1}
\begin{tabular}{|l|c|c|}
\hline
\textbf{Layer} & \textbf{Output Dimensions} & \textbf{Learnable Parameters} \\
\hline
Input & $2 \times \beta$ & 0 \\
1st 1D Conv  & $32 \times (\beta-2)$ & $224$ \\
1st BN  & $32 \times (\beta-2)$ & $64$ \\
2nd 1D Conv & $32 \times (\beta-7)$ & $6176$ \\
2nd BN  & $32 \times (\beta-7)$ & $64$ \\
MH-SA & $32 \times (\beta-7)$ & $ 512 \times(\beta-7) +49,152$ \\
GAP & $32$ & 0 \\
1st FCL & 64 & $2112$ \\
2nd FCL & 2 & $130$ \\
\hline
\end{tabular}
\end{table}

\subsubsection{Training Stage \texorpdfstring{$E_b/N_0$}{Eb/N0}}

The energy-per-bit to noise-power-spectral-density ratio ($E_b/N_0$) during the training phase constitutes a critical parameter that significantly impacts the generalization capability of the DNN-based demodulator. An excessively high $E_b/N_0$ value may cause the network to overfit to noise-free conditions, resulting in poor performance in practical noise-impaired channels. Conversely, training with excessively low $E_b/N_0$ values may prevent the network from learning discriminative feature representations \cite{hyper-parameter-2}.

To determine the optimal training $E_b/N_0$ range, we employed Bayesian optimization methodology \cite{hyper-parameter-3}. The optimization objective was formulated to minimize the aggregate BER performance across the operational $E_b/N_0 \in [0, 30]$ dB. Through evaluation of candidate configurations, we identified an optimal training $E_b/N_0$ range of [24, 28] dB for the proposed DL-NOMA-CSK system. This range provides the DNN with sufficiently diverse channel conditions to learn robust feature representations while maintaining an adequate signal-to-noise ratio for effective pattern recognition.

\subsubsection{Loss Function and Optimization Algorithm}

For the training process, we employ binary cross-entropy loss, which is expressed as:
\begin{equation}
    \text{loss} = -\frac{1}{B}\sum_{j=1}^{B}[b_j\log(\hat{p}_j) + (1-b_j)\log(1-\hat{p}_j)],
\end{equation}
where $B$ is the mini-batch size, $b_j \in \{0,1\}$ denotes the true transmitted bit for the $j$-th sample, and $\hat{p}_j \in [0,1]$ is the predicted probability that $b_j$. This loss function is particularly effective for binary classification tasks and provides well-calibrated probability estimates, which are essential for reliable demodulation in the proposed system \cite{hyper-parameter-4}.

We employ the Adam optimizer with an initial learning rate of $10^{-3}$, which provides adaptive learning rate adjustments based on gradient moments for efficient convergence in chaotic signal demodulation. A learning rate scheduler reduces the learning rate by a factor of 0.1 when validation performance plateaus for 3 consecutive epochs. The model uses a mini-batch size of 64, balancing computational efficiency and gradient estimation quality.

Table \ref{tab:hyperparameters} summarizes the key hyperparameters employed in our final DNN-based demodulator configuration. This carefully tuned parameter set enables the network to effectively capture the intricate patterns in chaotic signals while maintaining robustness against channel impairments and interference in wireless communication scenarios.

\begin{table}[t]
\centering
\renewcommand\arraystretch{1.2}
\caption{\custsection{Hyperparameters Configuration}}
\label{tab:hyperparameters}
\begin{tabular}{lc}
\hline
Hyperparameters & Value \\
\hline
Convolutional Filters ($n$) & 32 \\
Kernel Size ($k_{size}$)& 3 \\
Attention Heads ($h$)& 8\\
Attention Dimension ($d_h$) &64\\
Loss Function & Binary Cross-Entropy \\
Optimization Algorithm & Adam \\
Initial Learning Rate & $10^{-3}$ \\
Mini-batch Size & 64 \\
Maximum Epochs & 20 \\
Training Stage $E_b/N_0$  & [24, 28] dB \\
\hline
\end{tabular}
\end{table}

\section{Performance Analysis}
\label{4}

In this section, we analyze and compare the computational complexity, EE, SE, and security performance of the proposed system with the benchmark traditional MU-OFDM-DCSK system \cite{MU-DCSK.1}, the SCMA-based SCS-MC-DCSK system \cite{SCMA-DCSK.1}, and the DL-assisted DL-SCMA-DCSK system \cite{DL-SCMA-DCSK}.

\subsection{Complexity Analysis}

We analyze the computational complexity of the proposed DL-NOMA-CSK system following \cite{DL-SCMA-DCSK}, defined as the total number of operations per symbol at the receiver, expressed in terms of the asymptotic upper bound $\mathcal{O}(\cdot)$. Since DNN training is conducted offline, we evaluate only the online deployment complexity.

For each SIC stage, FFT-based PSD computation contributes $\mathcal{O}(\beta \log_2 \beta)$. The DNN demodulator comprises 1D Conv layers with $\mathcal{O}(nk_{\text{size}}\beta)$ complexity, MH-SA mechanisms with dominant complexity $\mathcal{O}(nhd_h\beta + n\beta^2)$, and GAP/FCL layers with $\mathcal{O}(n\beta + n)$.

\begin{table*}[!t]
\centering
\caption{\custsection{Complexity, EE and SE Comparison Among the Proposed DL-NOMA-CSK, MU-OFDM-DCSK, SCS-MC-DCSK, and DL-SCMA-DCSK}}
\label{tab:performance_comparison}
\renewcommand{\arraystretch}{1}
\resizebox{0.75\textwidth}{!}{%
\begin{threeparttable}
\begin{tabular}{|l|c|c|c|}
\hline
\textbf{System} & \textbf{Complexity $\mathcal{O}(\cdot)$} & \textbf{EE $\eta$} & \textbf{SE $\xi$}\\
\hline
MU-OFDM-DCSK \cite{MU-DCSK.1} & $\mathcal{O}((K+1-N)N\beta)$ & $\frac{K-N}{K}$ & $\frac{(K-N)N}{\beta K}$ \\
\hline
SCS-MC-DCSK \cite{SCMA-DCSK.1} & $\mathcal{O}(M^N)$ & $\frac{N/K(K-1)}{N/K(K-1)+1}$ & $\frac{N/K(K-1)\log_2M}{\beta K}$ \\
\hline
DL-SCMA-DCSK \cite{DL-SCMA-DCSK} & $\mathcal{O}(n \cdot d_f \cdot K + (1+K)\beta)$ & $\frac{N}{N+1}$ & $\frac{N\log_2 M}{\beta K}$ \\
\hline
\textbf{DL-NOMA-CSK} & $\mathcal{O}(\beta^2 + \beta \log_2 \beta)$ & 1 & $\frac{N}{\beta}$ \\
\hline
\end{tabular}

\begin{tablenotes}
    \footnotesize
\item[1] $K$ is the number of subcarriers, $N$ is the number of users, $M$ is the codebook size, $d_f$ denotes the number of overlapping layers per subcarrier.
\end{tablenotes}
\end{threeparttable}
}
\end{table*}

The total complexity per symbol is $\mathcal{O}(\beta \log_2 \beta + nk_{\text{size}}\beta + nhd_h\beta + n\beta^2 + n\beta + n)$, dominated by $\mathcal{O}(n\beta^2 + nhd_h\beta)$. With $\beta = 128$ and hyperparameters from Table~\ref{tab:hyperparameters}, approximately 0.52M operations per symbol are required.

As shown in Table~\ref{tab:performance_comparison}, the proposed system exhibits higher complexity than MU-OFDM-DCSK \cite{MU-DCSK.1} and SCS-MC-DCSK \cite{SCMA-DCSK.1} systems. However, it avoids the exponential complexity growth inherent in systems where complexity scales with codebook size and user numbers, such as $\mathcal{O}(M^N)$ in SCS-MC-DCSK. Notably, its complexity is comparable to DL-SCMA-DCSK \cite{DL-SCMA-DCSK}, indicating that the computational overhead primarily stems from DL components rather than the underlying modulation scheme.

The proposed DNN architecture is highly parallelizable and can leverage the extensive parallel computing capabilities of modern processors. For BS equipped with substantial computational resources, this complexity level remains acceptable.

\subsection{Energy Efficiency Analysis}

We evaluate the EE of the proposed DL-NOMA-CSK system following the definition in \cite{SCMA-DCSK.1,DL-SCMA-DCSK}. The EE is defined as:
\begin{equation}
\eta = \frac{E_{\text{info}}}{E_b},
\end{equation}
where $E_{\text{info}}$ is the energy of the information-bearing signal and $E_b$ represents the energy per transmitted bit. The bit energy satisfies $E_b = E_{\text{ref}} + E_{\text{info}}$, where $E_{\text{ref}}$ is the energy of the reference signal.

In the proposed DL-NOMA-CSK system, the DNN-based demodulation eliminates reference signal transmission requirements during the online deployment phase by learning the intrinsic chaotic signal characteristics during offline training. Consequently, $E_{\text{ref}} = 0$, leading to an optimal EE of $\eta = 1$.

As shown in Table~\ref{tab:performance_comparison}, the proposed DL-NOMA-CSK system achieves superior EE compared to all baseline systems. This improvement becomes more pronounced as the number of users increases, since traditional schemes require proportionally more reference signals. The elimination of reference signal transmission not only improves EE but also reduces the overall power consumption, which is particularly beneficial for energy-constrained vehicular communication devices.

\subsection{Spectral Efficiency Analysis}

According to the definition of SE in \cite{MU-DCSK.2}, which is the ratio of the data rate to the total bandwidth, the SE of the proposed DL-NOMA-CSK scheme is derived as follows:
\begin{equation}
\xi = \frac{R_{\text{total}}}{B_c} = \frac{N/T_b}{1/T_c} = \frac{N}{\beta},
\end{equation}
where $\xi$ represents the SE, $N$ is the number of simultaneously transmitting vehicles, and $R_{\text{total}} = N/T_b$ is the total bit rate. The bandwidth $B_c = 1/T_c$ is determined by the chip rate under the ideal filtering assumption, with the relationship $T_b = \beta T_c$.

The proposed DL-NOMA-CSK system achieves superior SE compared to conventional MU-OFDM-DCSK and SCMA schemes. Traditional MU-DCSK systems require reference signal transmission and achieve an SE of approximately $\xi_{\text{MU-DCSK}} = {N_s}/{\beta}$, where $N_s < N$ represents the number of data subcarriers after allocating private subcarriers for reference signals. Similarly, SCMA-based schemes such as SCS-MC-DCSK and DL-SCMA-DCSK systems also exhibit reduced SE due to both reference signal requirements and fixed codebook constraints.

\subsection{Security Analysis}

We evaluate the security performance of the proposed DL-NOMA-CSK system in terms of the information leakage rate and the secrecy capacity. Assume that the probabilities of transmitting binary symbols "0" and "1" are equal for each vehicle. The mutual information between the transmitted data $\tilde{s}_i$ of the $i$-th vehicle and the data retrieved by the eavesdropper $R_E$ is calculated as \cite{hyper-parameter-2}:
\begin{align}
\label{MI}
&I_{E,i}(R_{E}; \tilde{s}_i) \nonumber \\
&= H(R_{E}) - H(R_{E}|\tilde{s}_i) \nonumber \\
&= 1 + \rho_{E,i}\log_2(\rho_{E,i}) + (1-\rho_{E,i})\log_2(1-\rho_{E,i}),
\end{align}
where $H(\cdot)$ represents the entropy operation, and $\rho_{E,i}$ denotes the BER of the $i$-th vehicle's transmission at the eavesdropping receiver.

The overall information leakage rate for the NOMA system with $N$ vehicles is then calculated as:
\begin{equation}
\Lambda = \frac{1}{N} \sum_{i=1}^{N} I_{E,i}(R_{E}; \tilde{s}_i).
\end{equation}

Based on the information leakage rate, the secrecy capacity of legitimate users can be derived as:
\begin{equation}
C_{\text{secrecy}} = \frac{1}{N} \sum_{i=1}^{N} I_{L,i}(R_{L}; \tilde{s}_i) - \Lambda,
\end{equation}
where $I_{L,i}$ denotes the mutual information between the transmitted data $\tilde{s}_i$ of the $i$-th vehicle and the data retrieved by the legitimate user $R_L$. The calculation of $I_{L,i}$ follows the same form as Eq.~(\ref{MI}), but with $\rho_{L,i}$ representing the BER of the $i$-th vehicle's transmission at the legitimate receiver.

The proposed DL-NOMA-CSK system provides enhanced security compared to conventional chaotic communication systems. Unlike traditional MU-DCSK schemes, our design eliminates the transmission of reference signals entirely, preventing eavesdroppers from exploiting reference information to perform brute-force attacks or reconstruct the chaotic maps. Additionally, the NOMA principle superimposes multiple vehicle signals in the power domain, creating a complex composite signal that significantly increases the difficulty for unauthorized users to separate and decode individual vehicle transmissions. The security performance advantages will be further validated through simulations in Section \ref{Simulation Results and Discussions}.

\section{Simulation Results and Discussions}
\label{Simulation Results and Discussions}

This section presents a comprehensive performance evaluation of the proposed DL-NOMA-CSK scheme. Specifically, we describe the simulation setup and dataset generation process, and compare the BER performance against benchmark schemes and other DL-based methods. Moreover, the security performance and robustness of our design are also analyzed.

\subsection{Simulation Settings and Dataset Discussion}
\label{Dataset Discussion}

\begin{table}[!t]
\caption{\custsection{V2I Channel Model Parameters}}
\label{table:channel_params}
\centering
\begin{tabular}{|l|c|c|}
\hline
\textbf{Parameter} & \textbf{Primary road} & \textbf{Auxiliary road} \\
\hline
Center frequency ($f_c$) & \multicolumn{2}{c|}{3.35 GHz} \\
\hline
Sampling rate & \multicolumn{2}{c|}{100 MHz} \\
\hline
Vehicle speed & 95 km/h & 50 km/h \\
\hline
T-R distance range & \multicolumn{2}{c|}{250-650 m} \\
\hline
K-factor mean ($\mu_{K}$) & 9.56 dB & 4.22 dB \\
\hline
K-factor std ($\sigma_{K}$)& 4.58 dB & 4.96 dB \\
\hline
RMS delay spread ($\mu_{DS}$) & 76.1 ns & 238.8 ns \\
\hline
RMS Doppler spread ($\mu_{DPS}$) & 33.3 Hz & 35.4 Hz \\
\hline
RMS angular spread ($\mu_{AS}$) & $20.8^\circ$ & $36.5^\circ$ \\
\hline
Number of paths ($L_i$) &  5-8  & 10-16 \\
\hline
PDP model & \multicolumn{2}{c|}{Exponential} \\
\hline
PDP decay factor & 60.0 ns  & 190.0 ns \\
\hline
\end{tabular}
\end{table}

Considering the complexity and time-varying characteristics of realistic vehicular communication environments. Beyond Rayleigh fading, we consider urban vehicle-to-infrastructure (V2I) channels \cite{c1} incorporating Doppler effects and Rician fading:
\begin{equation}
h_i(t) = \sum_{l=1}^{L_i}\lambda_{i,l}e^{-j2\pi f_{D,i,l}t}\delta(t-\tau_{i,l}), \quad 0 \leq t \leq T_b,
\end{equation}
where $L_i$ represents the number of multipath components, $\tau_{i,l}$ and $f_{D,i,l}$ are the time delay and Doppler frequency shift of the $l$-th path for the $i$-th vehicle. The path gain $\lambda_{i,l}$ is decomposed into a deterministic line-of-sight component and a stochastic scattered component, expressed as:
\begin{equation}
\lambda_{i,l} = \sqrt{\frac{K}{K+1}} + \sqrt{\frac{1}{K+1}}(\psi_{i,l,1} + j\psi_{i,l,2}),
\end{equation}
where $\psi_{i,l,1}$ and $\psi_{i,l,2}$ are independent Gaussian random variables with zero mean and unit variance, and $K$ is the Rician $K$-factor.

We consider both primary road and auxiliary road scenarios in urban V2I channels. The channel parameters are listed in Table \ref{table:channel_params}.

The DNN-based demodulator $\mathcal{D}(\cdot)$ is trained over 50 epochs with 1,000,000 samples per epoch. The factors governing dataset generation include the number of vehicles $N$, the spreading factor $\beta$, the training stage $E_b/N_0$, the number of fading paths $L_i$, time delay $\tau_{i,l}$, and various channel realizations.

In the online deployment phase, we follow a similar process to evaluate the performance of the proposed DL-NOMA-CSK scheme. In addition, all simulations are conducted using MATLAB R2024b and the NVIDIA RTX 4060Ti with CUDA 12.0 for computation.

\begin{table}[!t]
\centering
\caption{\custsection{Multipath Channel Parameters for Different Vehicles}}
\label{tab:channel_params}
\renewcommand{\arraystretch}{1.3}
\begin{tabular}{|c|c|c|c|}
\hline
\textbf{Vehicle} & \textbf{Number of} & \textbf{Path Power Gain} & \textbf{Time Delay} \\
$V_i$ & \textbf{Paths $L_i$} &$E(|\lambda_{i,l}|^2)$ & $\tau_{i,l}$ ($T_c$) \\
\hline
$V_1$ & 2 & $[1/2, 1/2]$ & $[0, 2]$ \\
\hline
$V_2$ & 3 & $[4/7, 2/7, 1/7]$ & $[0, 2, 4]$\\
\hline
$V_3$ & 4 & $[4/9, 2/9, 2/9, 1/9]$ & $[0, 2, 3, 5]$ \\
\hline
$V_4$ & 4 & $[7/12, 2/12, 1/12, 1/12]$ & $[0, 2, 4, 6]$ \\
\hline
\end{tabular}
\end{table}

\begin{figure}[!t] 
		\centering
        \captionsetup{justification=raggedright,labelsep=period,singlelinecheck=off} 
		\includegraphics[width=0.75\columnwidth]{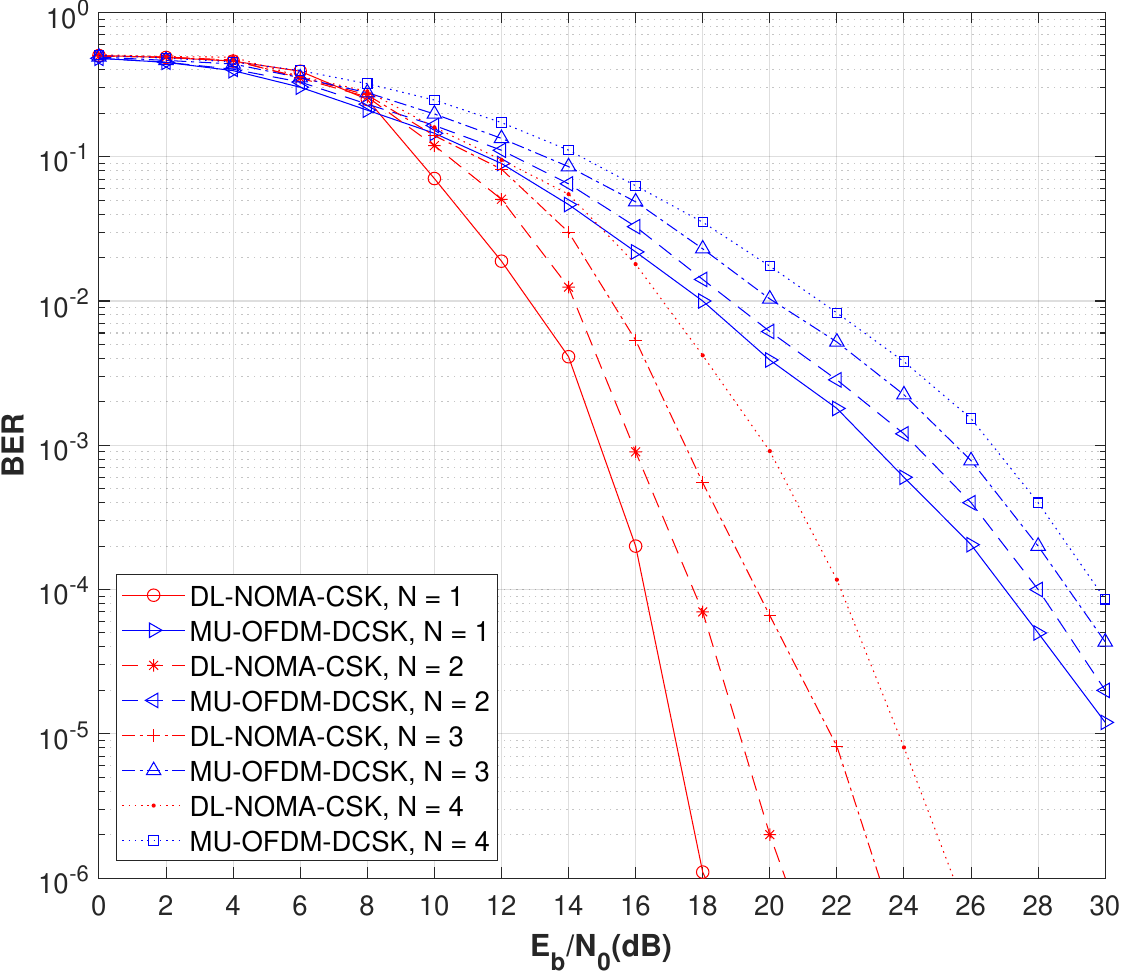 }
		\caption{BER performance comparisons with different number of vehicles $N$ between DL-NOMA-CSK and MU-OFDM- DCSK over multipath Rayleigh fading channel, where $\beta$ = 64.}
		\label{BER performance comparisons with different number of vehicles N between DL-NOMA-DCSK and MU-OFDM-DCSK over multipath Rayleigh fading channel}
	\end{figure}

\subsection{BER Performance Comparisons}

\subsubsection{BER Performance Comparisons With Benchmark Systems}

To evaluate the effectiveness of the proposed DL-NOMA-CSK scheme, we conduct BER performance comparisons against the benchmark MU-OFDM-DCSK system \cite{MU-DCSK.1}. In our simulations, both schemes utilize identical chaotic maps, with transmitted binary symbol "0" modulated by $\boldsymbol{\tilde{x}}$ and "1" by $\boldsymbol{\widehat{x}}$, and spreading factor $\beta = 64$ to ensure fair comparison. The multipath Rayleigh fading channel parameters for different vehicles $V_i$ are listed in Table \ref{tab:channel_params}, and MU-OFDM-DCSK allocates equal transmit power to all vehicles.

Fig. \ref{BER performance comparisons with different number of vehicles N between DL-NOMA-DCSK and MU-OFDM-DCSK over multipath Rayleigh fading channel} illustrates the BER performance comparison between the proposed DL-NOMA-CSK and MU-OFDM-DCSK schemes over multipath Rayleigh fading channels for different numbers of vehicles $N$. It can be seen that the proposed scheme consistently outperforms MU-OFDM-DCSK. This performance enhancement can be attributed to the synergistic combination of the NOMA power allocation strategy and DNN-based dual-domain demodulation, which jointly extract discriminative features from both time and frequency domains to effectively mitigate multi-user interference.

In addition, Fig. \ref{BER performance comparisons with different number of vehicles N between DL-NOMA-DCSK and MU-OFDM-DCSK over V2I channel, in (a) primary road and (b) auxiliary road scenarios.} demonstrates that the proposed system can also achieve better BER performance compared to the benchmark over V2I channels. It can be observed that for both Rayleigh fading and V2I channels, the BER performance of the proposed system degrades with increasing $N$. The reason is that NOMA is an interference-limited system. A practical solution is to partition vehicles with more distinctive channel conditions into groups, implement NOMA within each group, and allocate orthogonal resources across groups via OMA \cite{1-11}.

    \begin{figure}[!t]
\centering

\begin{subfigure}{0.75\columnwidth}
        \includegraphics[width=\columnwidth]{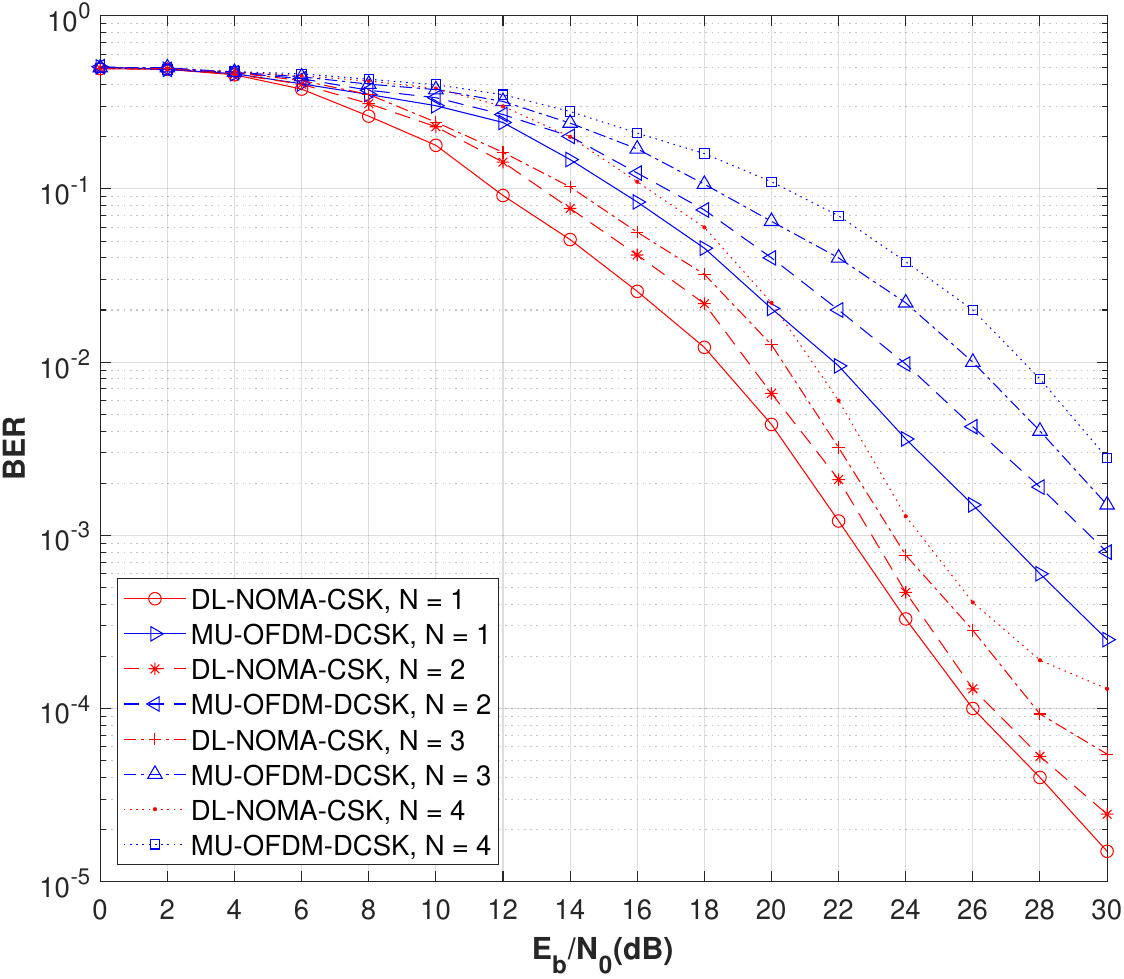}
        \caption{}
    \end{subfigure}

\begin{subfigure}{0.75\columnwidth}
        \includegraphics[width=\columnwidth]{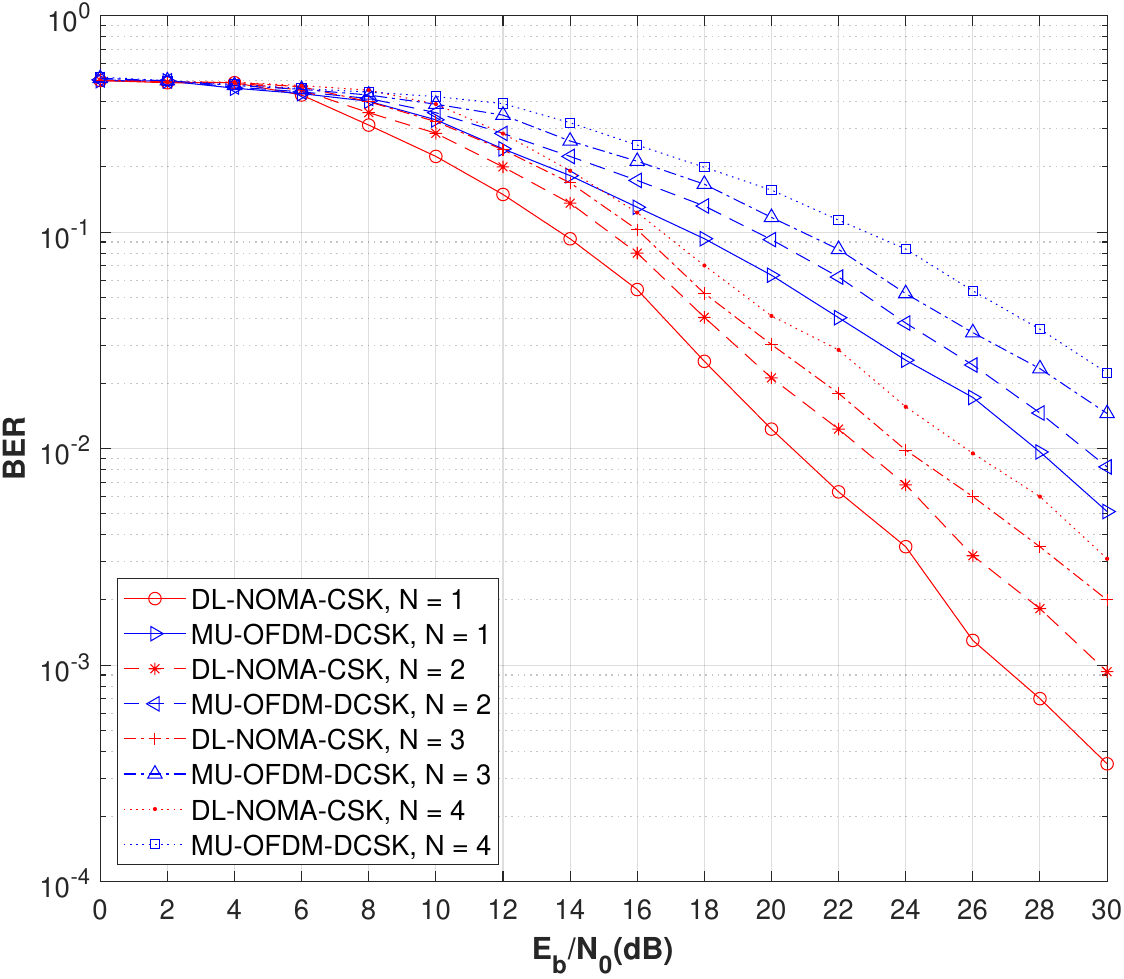}
       \caption{}
\end{subfigure}

\caption{BER performance comparisons with different numbers of vehicles $N$ between DL-NOMA-CSK and MU-OFDM-DCSK over V2I channels with $\beta = 64$, in (a) primary road and (b) auxiliary road scenarios.}
\label{BER performance comparisons with different number of vehicles N between DL-NOMA-DCSK and MU-OFDM-DCSK over V2I channel, in (a) primary road and (b) auxiliary road scenarios.}
\end{figure}

\subsubsection{BER Performance Comparisons With SCMA and Other DL-Assisted Systems}

We also compare the BER performance of the proposed DL-NOMA-CSK scheme with the SCMA-based SCS-MC-DCSK system \cite{SCMA-DCSK.1} and the other DL-assisted scheme, DL-SCMA-DCSK \cite{DL-SCMA-DCSK}. For fair comparison, although SCS-MC-DCSK and DL-SCMA-DCSK support multiple subcarriers, we configure them with a single information subcarrier to ensure fair evaluation with consistent system complexity and bandwidth utilization across all tested schemes.

\begin{figure}[!t] 
    \centering
    \captionsetup{justification=raggedright,labelsep=period,singlelinecheck=off}

    \includegraphics[width=0.75\columnwidth]{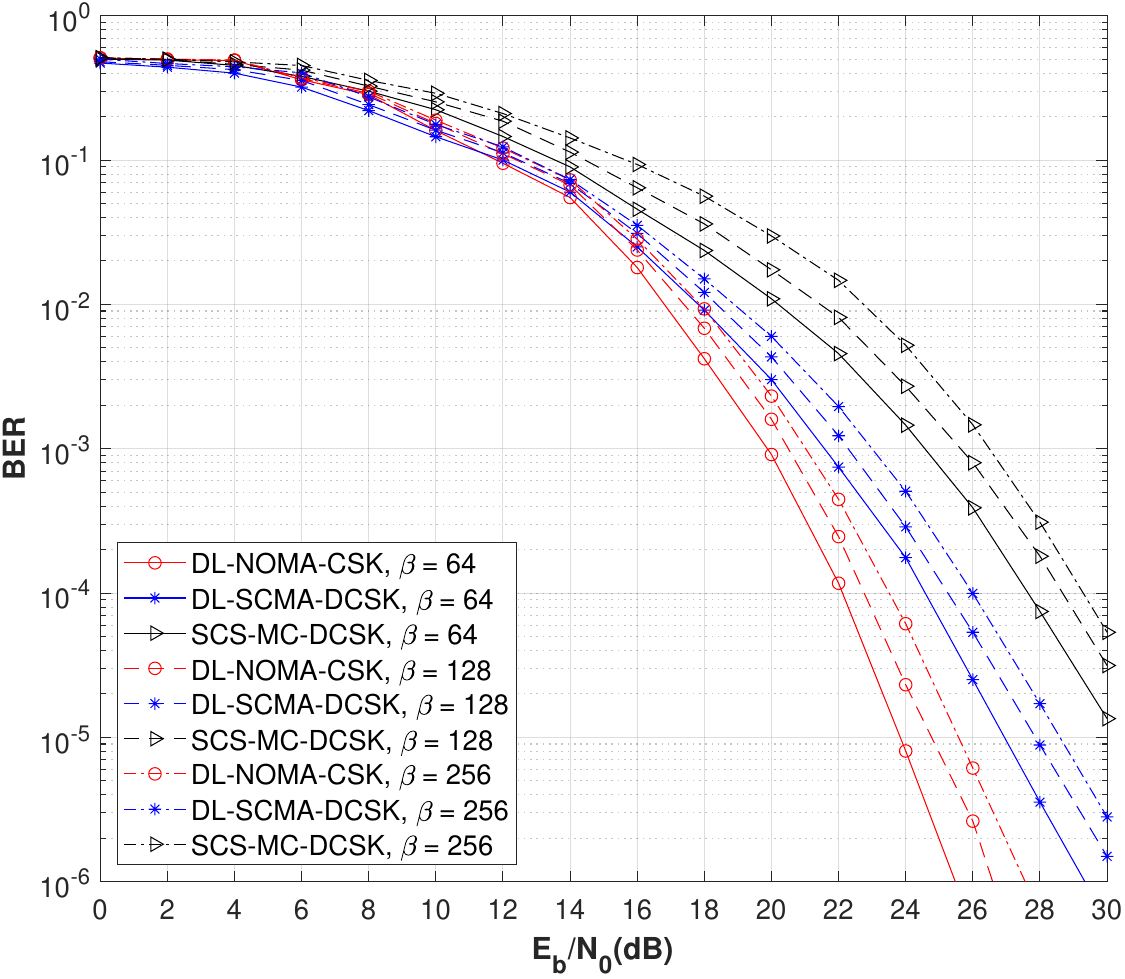} 
    \caption{BER performance comparisons with different values of $\beta$ between DL-NOMA-CSK, SCS-MC-DCSK, and DL-SCMA-DCSK over multipath Rayleigh fading channels with $N = 4$.}
    \label{BER performance comparisons with different number of beta between DL-NOMA-DCSK, SCS-MC-DCSK and DL-SCMA-DCSK over multipath Rayleigh fading channel.}
\end{figure}

    \begin{figure}[!t]
\centering

    \begin{subfigure}{0.75\columnwidth}
        \includegraphics[width=\columnwidth]{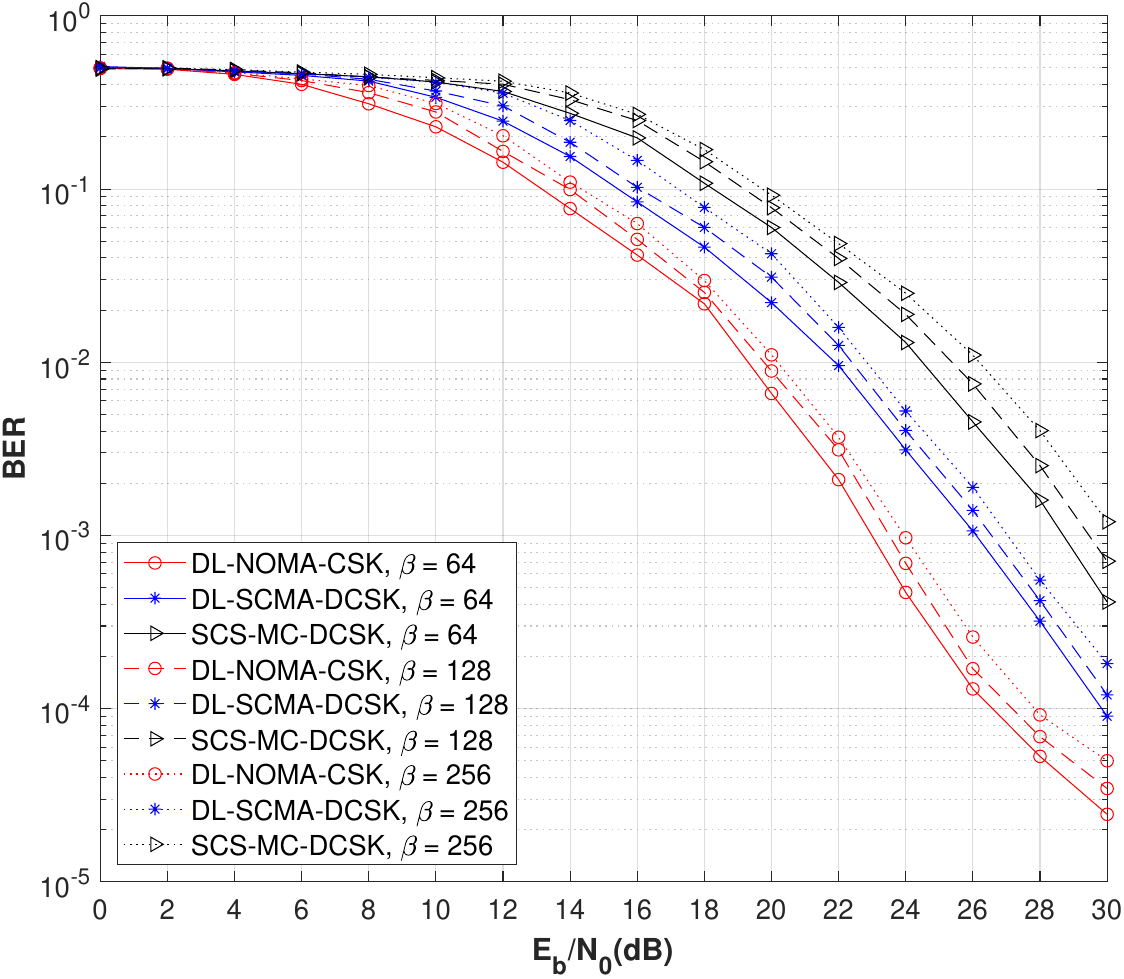}
        \caption{}
    \end{subfigure}

    \begin{subfigure}{0.75\columnwidth}
        \includegraphics[width=\columnwidth]{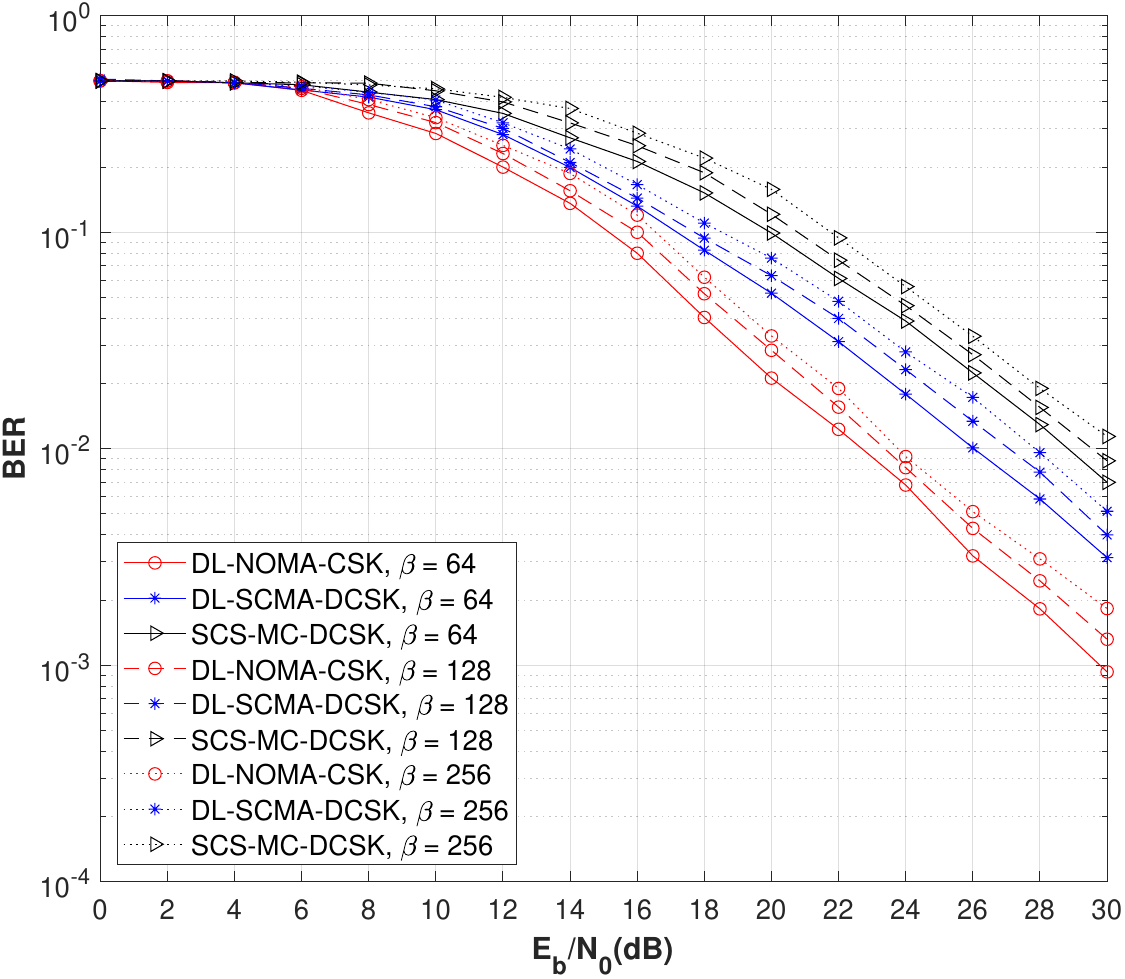}
        \caption{}
    \end{subfigure}
    
    \caption{BER performance comparisons with different values of $\beta$ between DL-NOMA-CSK, SCS-MC-DCSK, and DL-SCMA-DCSK over V2I channels with $N = 2$, in (a) primary road and (b) auxiliary road scenarios.}
    \label{BER performance comparisons with different number of beta between DL-NOMA-DCSK, SCS-MC-DCSK and DL-SCMA-DCSK over V2I channel, in (a) primary road and (b) auxiliary road scenarios.}
\end{figure}

Fig. \ref{BER performance comparisons with different number of beta between DL-NOMA-DCSK, SCS-MC-DCSK and DL-SCMA-DCSK over multipath Rayleigh fading channel.} and Fig. \ref{BER performance comparisons with different number of beta between DL-NOMA-DCSK, SCS-MC-DCSK and DL-SCMA-DCSK over V2I channel, in (a) primary road and (b) auxiliary road scenarios.} present comprehensive BER performance comparisons for different spreading factors $\beta$ over multipath Rayleigh fading and V2I channels, respectively. The simulation results reveal that as $\beta$ increases, all schemes experience performance degradation due to feature dilution and increased symbol duration, but the proposed system consistently outperforms other schemes. In V2I scenarios shown in Fig. \ref{BER performance comparisons with different number of beta between DL-NOMA-DCSK, SCS-MC-DCSK and DL-SCMA-DCSK over V2I channel, in (a) primary road and (b) auxiliary road scenarios.}, the performance advantage becomes more pronounced.

\subsection{Security Performance Analysis}
\label{Security Performance Analysis}

\begin{figure}[!t] 
		\centering
        \captionsetup{justification=raggedright,labelsep=period,singlelinecheck=off} 
		\includegraphics[width=0.75\columnwidth]{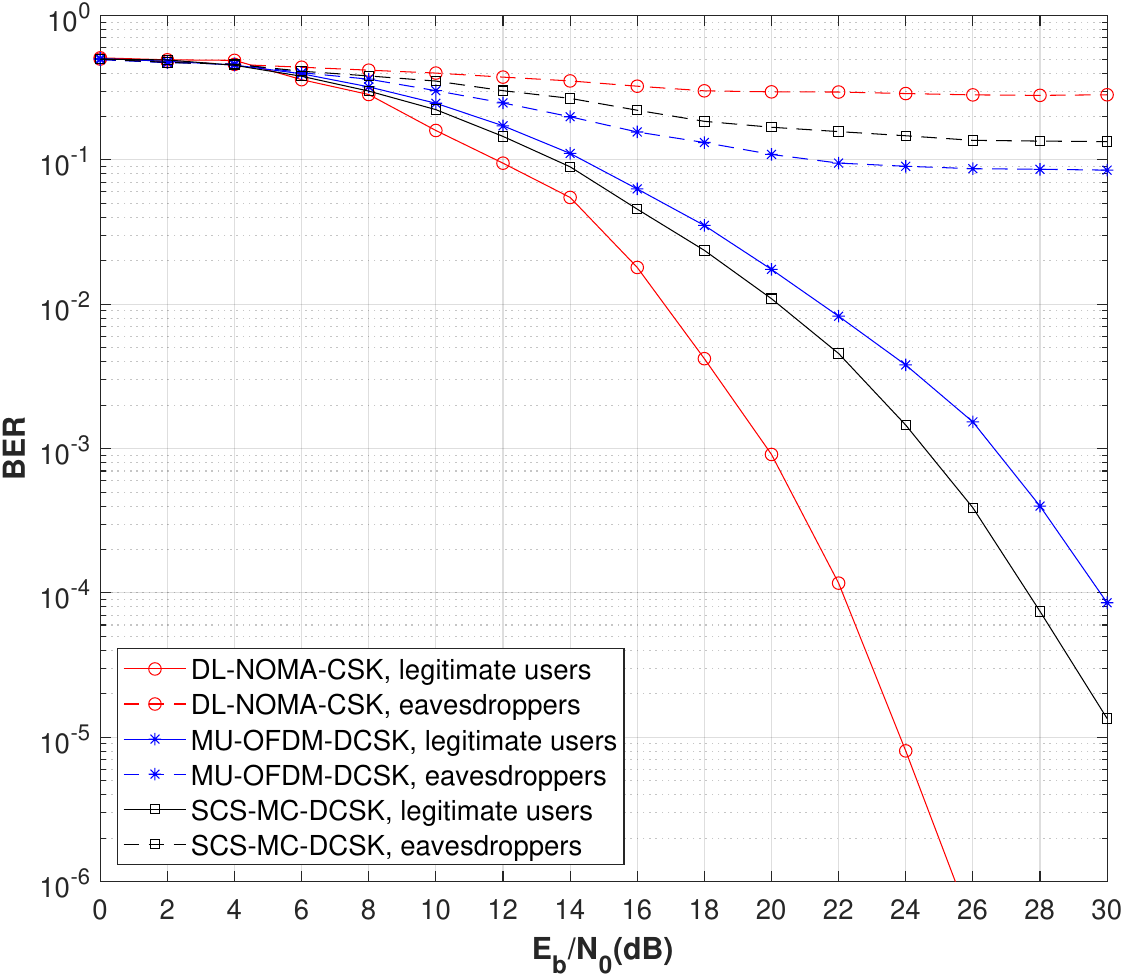} 
		\caption{BER performance comparisons of legitimate users and eavesdroppers between DL-NOMA-CSK, MU-OFDM- DCSK, and SCS-MC-DCSK over multipath Rayleigh fading channels with $\beta = 64$ and $N = 4$.}
		\label{BER performance comparisons of the legitimate users and the eavesdroppers between DL-NOMA-DCSK, MU-OFDM-DCSK and SCS-MC-DCSK over multipath Rayleigh fading channel.}
	\end{figure}

We investigate the BER performance of legitimate users and eavesdroppers in the proposed DL-NOMA-CSK scheme compared with MU-OFDM-DCSK \cite{MU-DCSK.1} and SCS-MC-DCSK \cite{SCMA-DCSK.1} over multipath Rayleigh fading channels. We assume that the eavesdropper can intercept the complete transmitted signals through the wireless channel, and employs an unsupervised learning approach \cite{Unsupervised} for signal demodulation. Unlike the legitimate BS that can train the DNN-based demodulator $\mathcal{D}(\cdot)$ using correctly labeled training data, the eavesdropper must first estimate the transmitted bit values from the intercepted signals and then utilize them as the training set. This process inevitably introduces cumulative errors that degrade the eavesdropper's demodulation performance.

        \begin{figure}[!t]
\centering
 \begin{subfigure}{0.75\columnwidth}
        \includegraphics[width=\columnwidth]{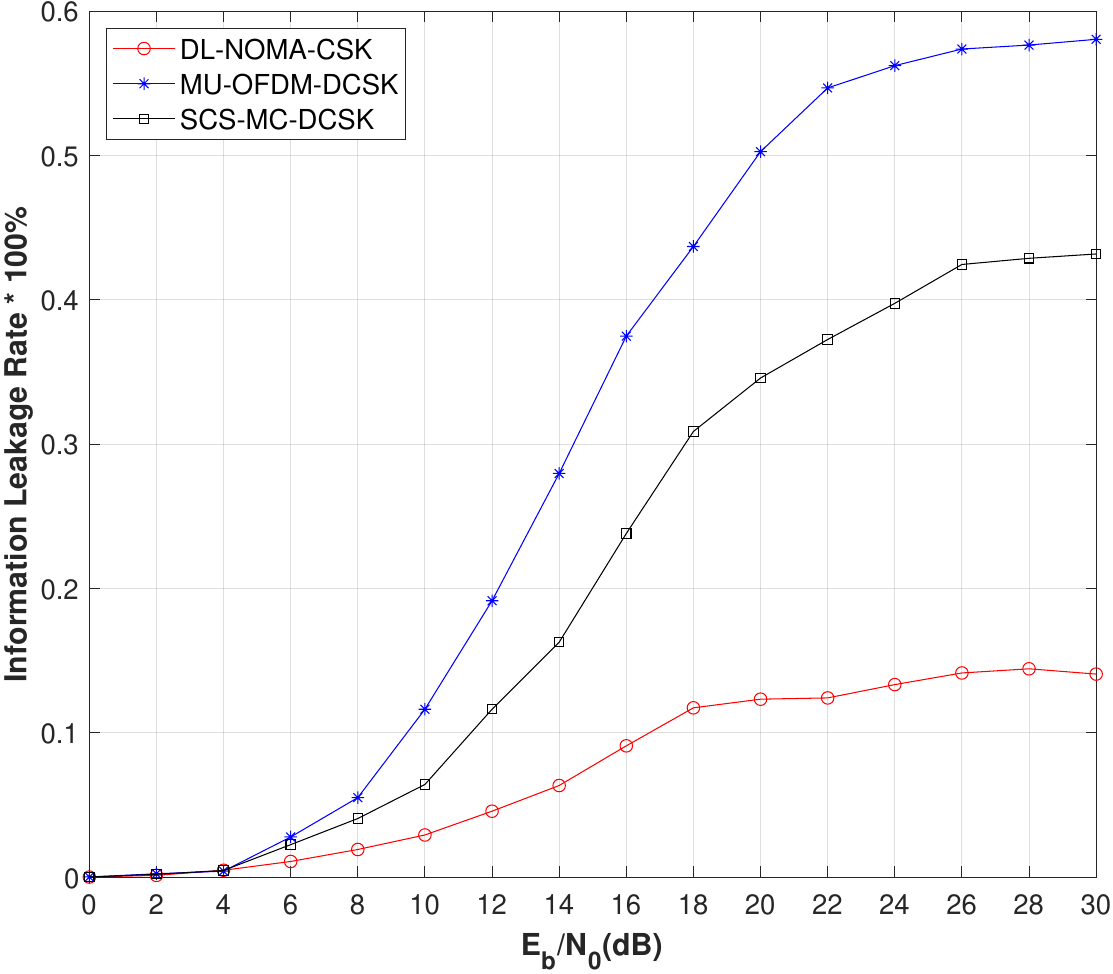}
        \caption{}
        \label{Information leakage rate}
    \end{subfigure}

\begin{subfigure}{0.75\columnwidth}
        \includegraphics[width=\columnwidth]{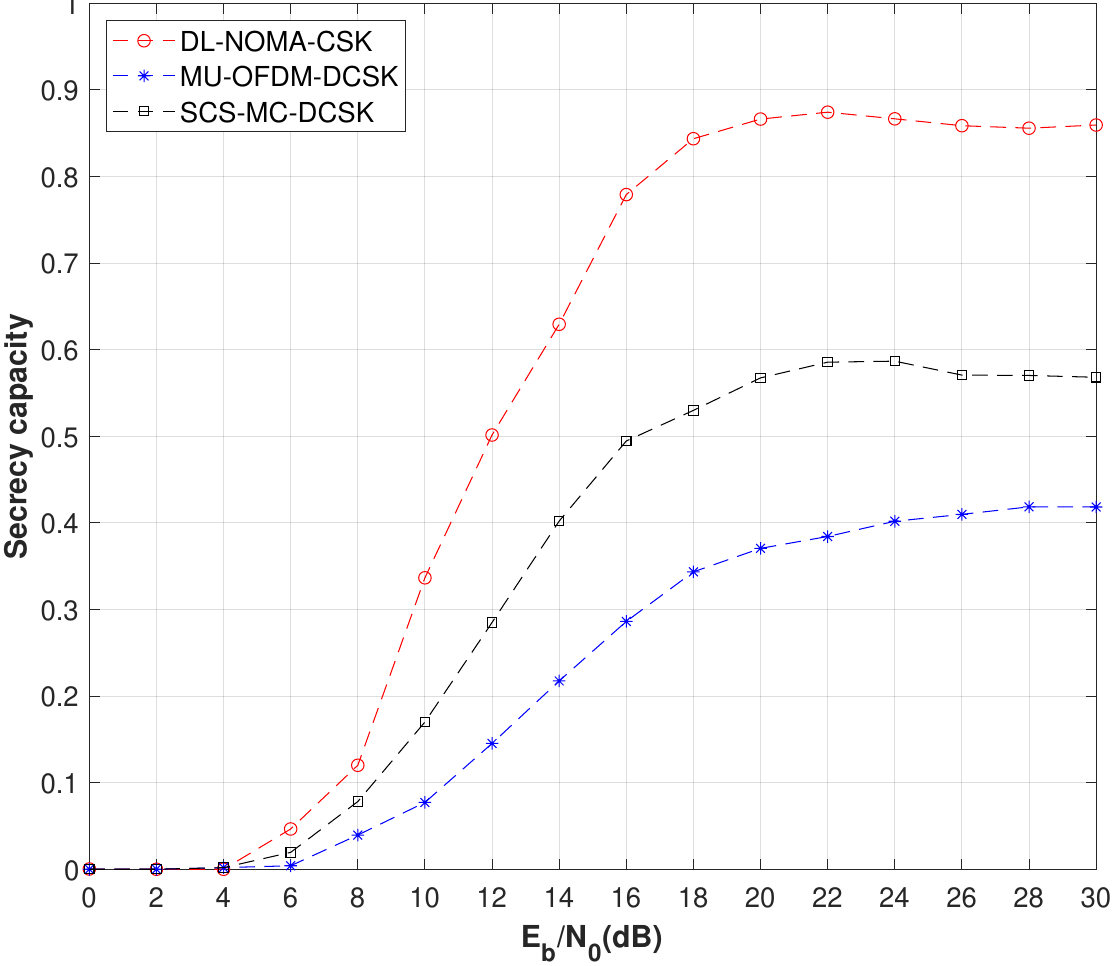}
       \caption{}
       \label{Secrecy capacity}
    \end{subfigure}

 \caption{Security performance comparisons between DL-NOMA-CSK, MU-OFDM-DCSK, and SCS-MC-DCSK over multipath Rayleigh fading channels with $\beta = 64$ and $N = 4$. (a) Information leakage rate. (b) Secrecy capacity.}
\label{Security performance comparisons between DL-NOMA-DCSK, MU-OFDM-DCSK and SCS-MC-DCSK over multipath Rayleigh fading channel.}
\end{figure}

It can be observed from Fig.~\ref{BER performance comparisons of the legitimate users and the eavesdroppers between DL-NOMA-DCSK, MU-OFDM-DCSK and SCS-MC-DCSK over multipath Rayleigh fading channel.} that all three systems exhibit significantly high BER for eavesdroppers, with DL-NOMA-CSK achieving the largest performance gap between legitimate users and eavesdroppers. This stems from chaotic signals' noise-like characteristics causing high cross-correlation among all transmitted signals \cite{r3-2}, impeding unsupervised learning convergence.  In the proposed DL-NOMA-CSK system, this difficulty is further exacerbated by the absence of reference signals, which eliminates the structural information that could potentially aid the eavesdropper's signal separation and demodulation process.

Next, we evaluate the secrecy capacity of legitimate users. As shown in Fig.~\ref{Security performance comparisons between DL-NOMA-DCSK, MU-OFDM-DCSK and SCS-MC-DCSK over multipath Rayleigh fading channel.}(a), the information leakage rate of MU-OFDM-DCSK adn SCS-MC-DCSK increases significantly with $E_b/N_0$. In contrast, the proposed DL-NOMA-CSK system maintains a remarkably low information leakage rate across all $E_b/N_0$ values, demonstrating that the elimination of reference signals effectively prevents eavesdroppers from decoding the information-bearing chaotic sequences.

Fig.~\ref{Security performance comparisons between DL-NOMA-DCSK, MU-OFDM-DCSK and SCS-MC-DCSK over multipath Rayleigh fading channel.}(b) illustrates the secrecy capacity performance. For MU-OFDM-DCSK and SCS-MC-DCSK, the secrecy capacity saturates after $E_b/N_0 = 18$ dB due to the aggravated information leakage. Notably, the proposed DL-NOMA-CSK system achieves superior secrecy capacity exceeding 0.9 at high $E_b/N_0$, benefiting from both the low information leakage rate and excellent BER performance of legitimate users. These results validate that the proposed system effectively enhances the secure information rate and provides robust physical layer security.

    \begin{figure}[!t]
\centering
 \begin{subfigure}{0.75\columnwidth}
        \includegraphics[width=\columnwidth]{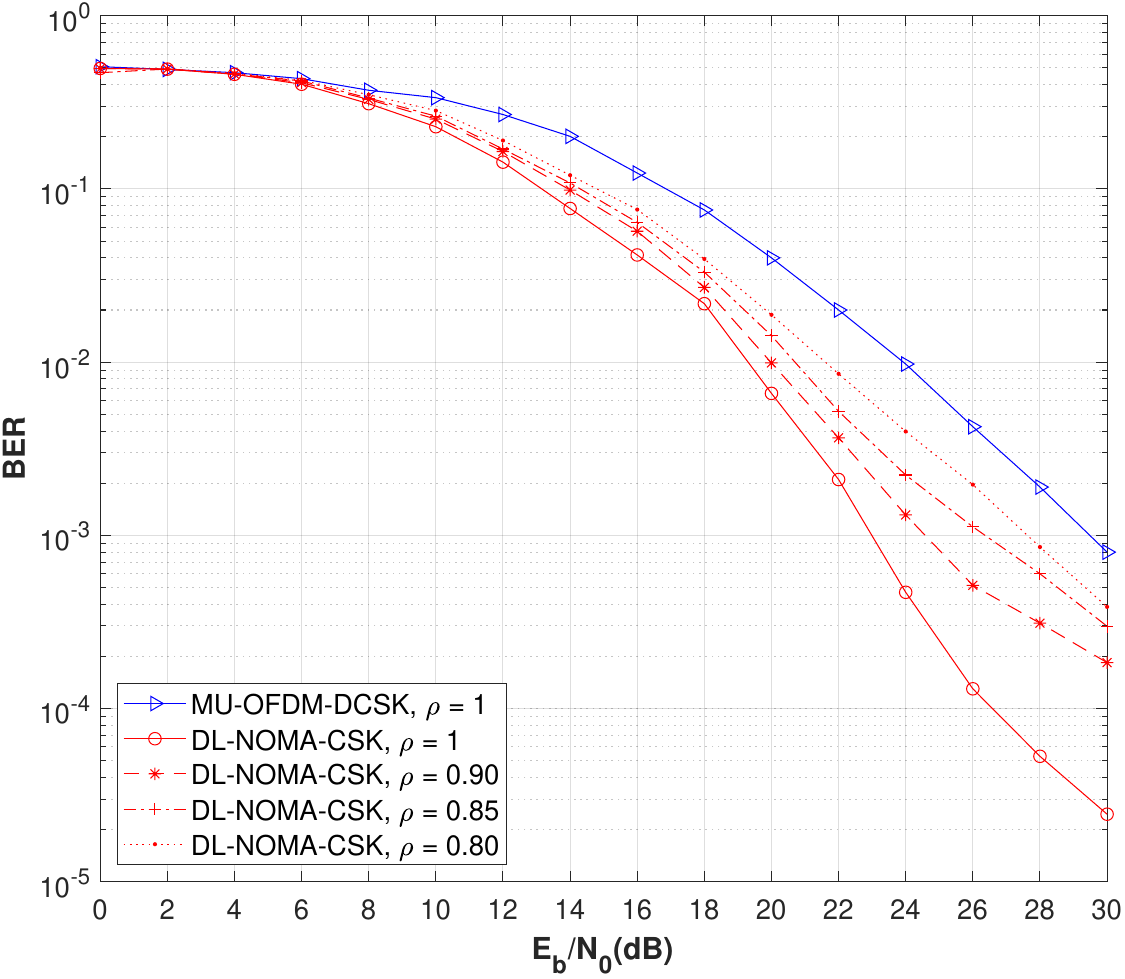}
        \caption{}
    \end{subfigure}

\begin{subfigure}{0.75\columnwidth}
        \includegraphics[width=\columnwidth]{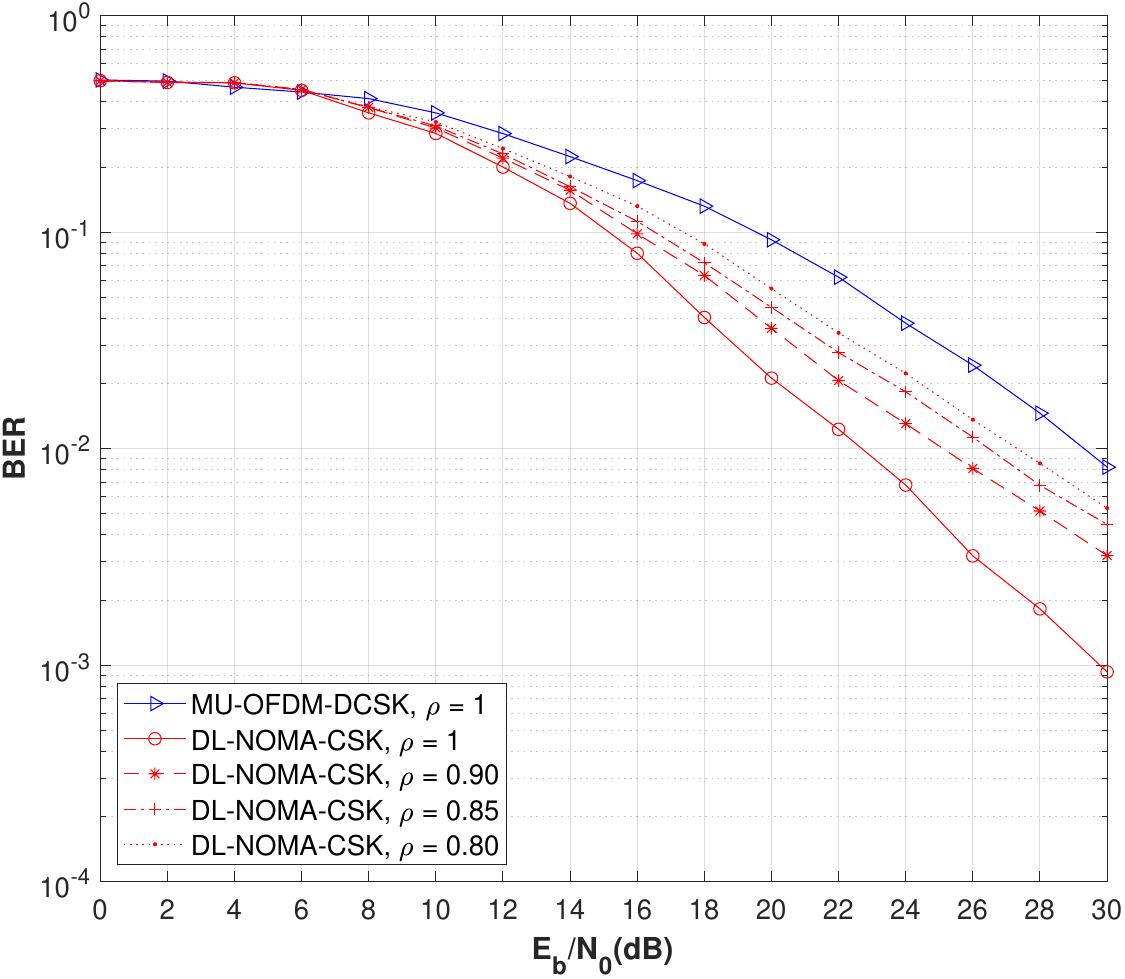}
       \caption{}
    \end{subfigure}

 \caption{BER performance comparisons with imperfect CSI between DL-NOMA-CSK and MU-OFDM-DCSK over V2I channel, where $\beta$ = 64 and $N$ = 2, in (a) primary road and (b) auxiliary road scenarios.}
\label{BER performance comparisons with imperfect CSI between DL-NOMA-DCSK and MU-OFDM-DCSK over V2I channel.}
\end{figure}

\subsection{Robustness Analysis}
\label{Robustness Analysis}

In the simulations presented above, perfect CSI is assumed at both the offline DNN training stage and the online deployment stage. However, in practical vehicular communication systems, CSI is typically imperfect due to channel estimation errors and the time-varying nature of wireless channels. To evaluate the generalization capability and practical applicability of the proposed DL-NOMA-CSK scheme, this section investigates the impact of imperfect CSI by analyzing the BER performance under channel conditions that deviate from the training assumptions. The imperfect CSI can be modeled using the correlation model \cite{CSI}:
\begin{equation}
    \widehat{h}_i = \rho h_i + \sqrt{1-\rho^2}\xi_i,
\end{equation}
where $h_i$ represents the equivalent channel gain that captures the combined effect of multipath fading over the bit duration $T_b$ for vehicle $V_i$, $\widehat{h}_i$ denotes the estimated channel coefficient, $\rho \in [0,1]$ characterizes the channel estimation accuracy, and $\xi_i \sim \mathcal{CN}(0,1)$ is the estimation error uncorrelated with $h_i$.

For time-varying V2I channels, the theoretical relationship is given by $\rho = J_0(2\pi f_D \tau)$, where $J_0(\cdot)$ is the zeroth-order Bessel function of the first kind, $f_D$ is the maximum Doppler frequency, and $\tau$ represents the CSI feedback delay. In this work, we directly vary $\rho$ to evaluate system robustness under different CSI quality levels. When $\rho = 1.0$, perfect CSI is achieved, representing the ideal case where training and deployment conditions match perfectly. As $\rho$ decreases, the CSI quality degrades, and the mismatch between training and deployment increases.

It can be observed from Fig.~\ref{BER performance comparisons with imperfect CSI between DL-NOMA-DCSK and MU-OFDM-DCSK over V2I channel.} that under V2I channel conditions, the proposed DL-NOMA-CSK system consistently outperforms the MU-OFDM-DCSK system regardless of the value of $\rho$ in both primary road and auxiliary road scenarios. However, as $\rho$ decreases, the performance gain achieved by the DL-NOMA-CSK system gradually diminishes. This is because the transmission characteristics learned during training do not match the channel conditions during deployment under imperfect CSI, causing the DNN-based demodulator to exhibit degraded feature extraction performance. Nevertheless, even under severe CSI imperfection ($\rho = 0.85$), the proposed scheme still demonstrates better BER performance under V2I channel conditions, indicating satisfactory robustness and generalization capability.

\section{Conclusion}
\label{6}

This paper proposes a novel DL-NOMA-CSK scheme to tackle the critical challenges of secure and efficient MU transmission in vehicular communication systems by integrating DL-assisted demodulation with PD-NOMA. Unlike conventional MU-DCSK systems with low spectral efficiency and SCMA-based solutions with high computational complexity, the proposed design effectively balances performance, complexity, and security. To achieve this, the proposed system eliminates reference signal transmission by employing a DNN-based demodulator that learns the intrinsic characteristics of chaotic signals. By jointly processing time-domain and frequency-domain features, the DNN-based demodulator extracts discriminative characteristics of chaotic signals and achieves enhanced robustness under dynamic vehicular channel conditions. Moreover, by integrating the DNN into the SIC framework, the proposed system effectively mitigates error propagation issues inherent in conventional NOMA receivers. In summary, the proposed DL-NOMA-CSK scheme provides a promising solution for next-generation vehicular communication systems requiring enhanced physical layer security, massive connectivity, and efficient spectrum utilization. Future work will investigate multicarrier extensions, integration with advanced channel estimation techniques, and practical deployment optimization.

\bibliographystyle{IEEEtran}
\bibliography{references}

\end{document}